\documentclass[a4paper,11pt]{article}

\usepackage{jheppub} 

\usepackage[T1]{fontenc} 
\usepackage{amssymb, bm, graphicx, graphics, mathrsfs, amsmath}
\usepackage{subfigure} \subfiguretopcaptrue

\newcommand{\ben}{\begin{eqnarray}}
\newcommand{\een}{\end{eqnarray}}

\newcommand{\bp}{\bar\phi}
\newcommand{\Pm}{_{;\mu}}
\newcommand{\Pn}{_{;\nu}}
\newcommand{\Pu}{_{,u}}
\newcommand{\Pv}{_{,v}}
\newcommand{\Puv}{_{,uv}}

\newcommand{\bz}{\bar{z}}

\newcommand{\bw}{\bar{w}}

\makeatletter
\newcommand{\raisemath}[1]{\mathpalette{\raisem@th{#1}}}
\newcommand{\raisem@th}[3]{\raisebox{#1}{$#2#3$}}
\makeatother

\DeclareRobustCommand{\rchi}{{\mathpalette\irchi\relax}}
\newcommand{\irchi}[2]{\raisebox{\depth}{$#1\chi$}} 

\title{\boldmath Scalar field as~an~intrinsic time measure in~coupled dynamical matter-geometry systems.\\ I. Neutral gravitational collapse}

\author[a,b]{Anna Nakonieczna,}
\author[c]{and Dong-han Yeom}

\affiliation[a]{Institute of~Physics, Maria Curie-Sk{\l}odowska University, \\
Plac Marii Curie-Sk{\l}odowskiej 1, 20-031 Lublin, Poland}
\affiliation[b]{Institute of~Agrophysics, Polish Academy of~Sciences,\\
Do{\'s}wiadczalna 4, 20-290 Lublin, Poland}
\affiliation[c]{Leung Center for Cosmology and~Particle Astrophysics,\\
National Taiwan University, Taipei 10617, Taiwan}

\emailAdd{aborkow@kft.umcs.lublin.pl}
\emailAdd{innocent.yeom@gmail.com}

\abstract{
There does not exist a~notion of~time which could be transferred straightforwardly from classical to~quantum gravity. For~this reason, a~method of~time quantification which would be appropriate for gravity quantization is being sought. One of~the~existing proposals is using the~evolving matter as~an~intrinsic `clock' while investigating the~dynamics of~gravitational systems. The~objective of~our~research was to~check whether scalar fields can serve as~time variables during a~dynamical evolution of~a~coupled multi-component matter-geometry system. We~concentrated on~a~neutral case, which means that the~elaborated system was not charged electrically nor magnetically. For~this purpose, we investigated a~gravitational collapse of~a~self-interacting complex and~real scalar fields in~the Brans-Dicke theory using the~2+2 spacetime foliation. We~focused mainly on~the region of~high curvature appearing nearby the~emerging singularity, which is essential from the~perspective of~quantum gravity. We~investigated several formulations of~the~theory for various values of~the~Brans-Dicke coupling constant and~the~coupling between the~Brans-Dicke field and~the~matter sector of~the~theory. The~obtained results indicated that the~evolving scalar fields can be treated as~time variables in~close proximity of~the~singularity due to~the~following reasons. The~constancy hypersurfaces of~the~Brans-Dicke field are spacelike in~the vicinity of~the~singularity apart from the~case, in~which the~equation of~motion of~the~field reduces to~the~wave equation due to~a~specific choice of~free evolution parameters. The~hypersurfaces of~constant complex and~real scalar fields are spacelike in~the~regions nearby the~singularities formed during the~examined process. The~values of~the~field functions change monotonically in~the~areas, in~which the~constancy hypersurfaces are spacelike.
}

\begin{document} 
\maketitle
\flushbottom

\section{Introduction}
\label{sec:intro}

The~problem of~time quantification is vital mainly in~canonical approaches to~quantum gravity, because in~gravitational systems there does not exist a~notion of~time which could be straightforwardly transferred between the~classical and~quantum levels. The~issue of~measuring time is primarily significant in~investigations of~the~dynamics of~quantized gravitational systems. Moreover, seeking alternative descriptions of~a~gravitational system temporal evolution may become useful also in~classical gravity, as~they can potentially facilitate examining dynamics of~complicated coupled matter-geometry systems.

The~non-standard approach to~measuring time during a~process proceeding due to~gravitational interaction is using the~evolving matter itself in~this regard~\cite{DeWitt1967}. Also geometry could be used as~a~`clock' when a~dynamical evolution of~matter is studied. In~general, the~dynamical behavior of~a~matter-geometry system can be followed with respect to~one of~its internal degrees of~freedom, which serves as~a~reference for~the~remaining degrees of~freedom and~is interpreted as~a~dynamical observer. In~order to~obtain a~successful description of~the~passage of~time, two conditions have to~be fulfilled during the~investigated part of~an evolution. First, the~selected spacetime slices, parametrized by a~time variable, ought to~be spacelike. Second, the~chosen time parametrization should remain monotonic during the~course of~the~evolution of~interest.

The~idea of~quantifying time described above has been widely employed in~diversified analyses related to~non-perturbative quantum gravity and~quantum cosmology, which involve investigating a~time evolution of~a~matter-geometry system without fixed geometry of~a~background spacetime~\cite{Rovelli1991}. The~constancy hypersurfaces of~a~scalar field were the~spacelike slices parametrized by a~time variable used in~the~construction of~a~Hamiltonian, which governed gauge transformations between these slices and~thus described a~quantum evolution of~a~gravitational field~\cite{RovelliSmolin1994}. A~scalar field was also treated as~a~time variable for~the~relational Dirac observables, whose dynamics was traced with respect to~it~\cite{DomagalaGieselKaminskiLewandowski2010,LewandowskiDomagalaDziendzikowski2012}. The~inflation epoch of~the~Universe was examined in~the~regime of~loop quantum gravity using a~scalar field with an~arbitrary potential with the~gauge for~the~Hamiltonian constraints, which ensured that the~constancy hypersurfaces of~the~scalar field were spacelike~\cite{AlexanderMaleckiSmolin2004}. A~similar approach involving~a~massless scalar field was used in~loop quantum cosmology~\cite{AshtekarPawlowskiSingh2006PRL,AshtekarPawlowskiSingh2006PRD}. The~tunneling decay rate~\cite{DabrowskiLarsen1995} of~a~simple harmonic universe~\cite{GrahamHornKachruRajendranTorroba2014} was calculated with the~use of~a~homogeneous, massless, minimally coupled scalar field additionally introduced to~the~model and~having a~negligible contribution to~the~total energy density of~the~system~\cite{MithaniVilenkin2012,MithaniVilenkin2015}. A~scalar field was treated as~a~common variable for~internal time and~a~Hamiltonian evolution parameter during constructing a~specific version of~the~Wheeler-DeWitt equation and~its quantum timelike counterpart~\cite{Perlov2015}. Apart from the~scalar field, dust and~radiative fluid were also used to~provide a~matter degree of~freedom, which allowed tracing the~temporal evolution of~a~gravitational system~\cite{BrownKuchar1995}.

The~non-minimally coupled to~geometry Brans-Dicke scalar field, which is a~part of~the~theoretical model considered in~the~present paper, was also analyzed in~the~context of~time measurements in~quantum cosmology~\cite{ZhangArtymowskiMa2013-084024}. Since the~field is a~monotonic function of~cosmological time, it~was assumed to~be a~good candidate for~an~internal time variable. The~cosmological Brans-Dicke model was quantized within the~framework of~loop quantum cosmology and~the~effective Hamiltonian was obtained under the~assumptions of~isotropy and~spatial flatness. Unitarity of~the~evolution in~the~Brans-Dicke quantum cosmological model with a~time variable in~the~form of~an~isotropic and~homogeneous matter fluid was elaborated in~\cite{AlmeidaBatistaFabrisMoniz2015}.

The~Brans-Dicke theories are straightforward extensions of~general relativity towards the~scalar-tensor theories of~gravity~\cite{BransDicke1961-925}. The~gravitational interaction is described within them by both a~scalar field and~the usual metric tensor of~the~Einstein theory. The~effective gravitational coupling changes within the~spacetime and~asymptotically attains the~value of~the gravitational constant $G$. The~strength of~the~coupling is determined by a~scalar field which asymptotically tends towards the~value of~$G^{-1}$. The~theory possesses the~so-called Brans-Dicke coupling, $\omega$, which is a~dimensionless constant. For~its small values, the~scalar constituent of~the~gravitational interaction predominates the~tensor component, while for its large values the~contribution of~the~tensorial part is more important. In~the~limit of~$\omega\to\infty$ the~Brans-Dicke theory becomes the~Einstein theory~\cite{Faraoni1999-084021,Will2014-4}. Several discussions on~the~problem whether this statement is correct in~general or~whether the~conclusion depends on~the~value of~the~trace of~the~stress-energy tensor of~the~theory, the~symmetry, staticity, stationarity or~asymptotic flatness of~the~solutions, can be found, e.g.,~in~\cite{RomeroBarros1993-243,Chauvineau2003-2617,ChauvineauSpallicciFournier2005-S457,Chauvineau2007-297}.

The~Brans-Dicke theories have been tested against the~experimental data. The~most convincing analyses were done within the~Solar System, as~the~considered theories are so close to~the~Einstein theory that they straightforwardly agree with all cosmological and~astrophysical observations, at~least for adequately big values of~$\omega$~\cite{Will2014-4}. The~frequency shift of~the~low energy photons due to~the~spacetime curvature experienced on~their road to~and~from the~Cassini spacecraft as~they passed nearby the~Sun confirmed the~appropriateness of~selected observational predictions of~the~Brans-Dicke theoretical construction~\cite{BertottiIessTortora2003-374}.

The~experimental restrictions on~the~value of~the~Brans-Dicke coupling can be posed on~the~basis of~a~series of~astronomical observations and~tests. The~above-mentioned Cassini-Huygens experiment gave a~lower bound on~$\omega$ equal to~$40000$~\cite{BertottiIessTortora2003-374,DeFeliceManganoSerpicoTrodden2006-103005,Will2014-4}. The~supernovae Ia data provided the~value of~$-1.477$ within the~Brans-Dicke cosmology with a~pressureless fluid~\cite{FabricGoncalvesRibeiro2006-49} and~$-1.9$ in~the~dimensionally reduced theory~\cite{Qiang2009-210}. When combined with the~Hubble parameter versus the~redshift relation measurements, information based on~the~Alcock-Paczy{\'n}ski test and~the baryon acoustic oscillations observational data, the~value of~$\omega$ was estimated as~$-0.8606$, $-1.1103$ and~$-2.3837$ for several dynamical setups of~the~Brans-Dicke cosmology in~the~vicinity of~the~de Sitter state~\cite{HrycynaSzydlowskiKamionka2014-124040}. The~cosmic microwave background radiation temperature measurements performed by Planck allowed constraining the~value of~$\omega$ to~be greater than $692$~\cite{Avilez2014-011101}. Their combination with the~polarization data obtained by WMAP and~the baryon acoustic oscillations distance ratio data from the~Sloan Digital Sky Survey and~the Six-degree-Field Galaxy Survey excluded the~range from $-407.0$ to~$175.87$ and~preferred the~values exceeding $181.65$~\cite{LiWuChen2013-084053}. According to~the~structure formation imprint in~the~cosmic background radiation, the~lower limit of~$\omega$ was equal to~$120$~\cite{AcquavivaEtAl2005-104025}. As~can be inferred from the~above collection of~diverse numerical data, the~experiment-based values of~the~Brans-Dicke coupling are strongly model-dependent, since various assumptions about the~symmetry and~matter content of~the~analyzed spacetime were made in~the~outlined analyses. Moreover, the~value of~$\omega$ may change as~the~Universe evolves. The~time scale of~the~dynamical process studied in~the~current paper, i.e.,~the~gravitational collapse, is negligible in~comparison to~the~cosmological time scales. For this reason, a~set of~constant values of~the~Brans-Dicke parameter was considered during the~performed investigations.

During the~cosmological evolution, the~scalar contribution to~gravity practically vanishes within most scalar-tensor formulations of~gravity~\cite{DamourNordtvedt1993-2219,DamourNordtvedt1993-3436,DamourPiazzaVeneziano2002-046007}. Hence, although $\omega$ is most probably large at~present and~for this reason the~Brans-Dicke theory is currently experimentally indistinguishable from the~Einstein theory, the~value of~the~coupling could have been smaller in~the~past. For~this reason, the~class of~the~Brans-Dicke theories is widely studied in~the~context of~the~early stages of~the~evolution of~the~Universe. Unlike general relativity, it~provides a~satisfactory mechanism of~the~transition between the~rapid inflationary stage of~the~Universe evolution and~its later cosmological phase, which is called the~extended inflation~\cite{LaSteinhardt1989-376,MathiazhaganJohri1998-29}.

The~corrections introduced by the~Brans-Dicke theory to~the~current values of~cosmological parameters such as~the~Hubble parameter, gravitational constant, or~the~usual fractions of~energy density within the~Friedman-Lema{\^\i}tre-Robertson-Walker cosmology, are negligible~\cite{ArikCalik2005-013,ArikCalikSheftel2008-225}. The~dimensionally reduced Brans-Dicke theory gave rise to~models of~an~accelerated expansion of~a~matter-dominated universe which are consistent with current observations and~with a~decelerating radiation-dominated epoch~\cite{SenSen2001-124006,QuiangMaHanYu2005-061501,deLeon2010-095002,deLeon2010-030}. The~late-time acceleration of~the~Universe expansion was also examined within the~Brans-Dicke theory under the~assumption of~the~spatial flatness of~the~Universe~\cite{Bisabr2012-427,Bisabr2012-127503} and~in the~presence of~a~fermionic field and~a~matter constituent described by the~barotropic equation of~state~\cite{SamojedenDevecchiKremer2010-027301,Liu2010-063523}. Cosmological implications of~holographic dark energy~\cite{NojiriOdintsov2006-1285,CapozzielloNojiriOdintsov2006-597,Setare2007-99,Setare2010-1} and~the stability of~agegraphic dark energy~\cite{FarajollahiSadeghiPouraliSalehi2012-79} were analyzed in~the~considered theory, while the~new holographic dark energy model was studied in~the~framework of~chameleon Brans-Dicke cosmology~\cite{ChattopadhyayPasquaKhurshudyan2014-3080}. The~isotropic and~homogeneous Brans-Dicke model with a~quartic scalar field potential and~barotropic matter explained the~accelerated expansion of~the~Universe without any assumptions about the~properties of~dark matter and~dark energy~\cite{HrycynaSzydlowski2013-016}. Constraints on~a~flat isotropic and~homogeneous Brans-Dicke cosmological model with matter in~the~form of~a~perfect fluid with a~constant equation of~state parameter were presented in~\cite{PaliathanasisTsamparlisBasilakosBarrow2015-arxiv}. The~evolution of~the~radius and~mass of~black holes in~an~expanding isotropic universe was discussed using the~Einstein-Straus model~\cite{EinsteinStraus1945-120} in~the~Brans-Dicke theory~\cite{SakaiBarrow2001-4717}. The~employed so-called `Swiss cheese' model described the~Friedman-Lema{\^\i}tre-Robertson-Walker universe, in~which spherical regions were replaced by Schwarzschild spacetimes.

For~finite values of~$\omega$, the~Brans-Dicke theory describes, under certain conditions, a~bouncing cosmology~\cite{Novello2008-127}, as~the scale factor does not vanish during the~backward temporal evolution of~the~Universe~\cite{FabrisFurtadoPeterPintoNeto2003-124003,TretyakovaShatskiyNovikovAlexeyev2012-124059,TretyakovaLatoshAlexeyev2015-185002}. Its value decreases to~a~minimum and~then increases, which allows avoiding the~presence of~the~initial singularity of~the~classical general relativistic cosmology. The~spatially flat and~isotropic cosmological model of~the~Brans-Dicke theory with $\omega\neq-1.5$ was quantized within the~loop quantum cosmology approach~\cite{ZhangArtymowskiMa2013-084024,ArtymowskiMaZhang2013-104010}. In~such a~setup of~the~effective loop quantum Brans-Dicke cosmology, the~classical initial singularity is replaced by a~quantum bounce.

The~researches on~using matter as~a~time variable described at~the~beginning of~this section did not address the~issue of~the~relevance of~such an~idea. The~behavior of~matter during the~investigated part of~an evolution and~within the~studied spacetime region was assumed to~be appropriate from the~perspective of~using it~as~a~`physical clock'. The~selected spacetime slices were presumed to~remain spacelike and~the~chosen time parametrization was presumed to~be monotonic during the~whole process. The~arguments supporting these assumptions are limited to~certain cases (e.g.,~the~homogeneity of~a~scalar field during the~inflation initial phase, which results in~a~spacelike character of~the~field constancy hypersurfaces~\cite{AlexanderMaleckiSmolin2004}). Hence, there exists a~need for research on~dynamical processes, which will justify or~contradict the~above-mentioned assumptions. An~introductory attempt in~respect of~validating these premises was made using the~simplest matter-geometry model, which involved a~gravitationally self-interacting scalar field minimally coupled to~gravity~\cite{NakoniecznaLewandowski2015-064031}. It~turned out that the~sole scalar field evolving in~the~spacetime under the~influence of~gravitational self-interaction possesses properties, which predestine it~for being a~time measurer, especially in~the~area of~high curvature nearby the~singularity. However, since the~examined field was the~only matter component in~the~spacetime, the~conducted studies are not relevant to~justify using the~scalar field as~a~`clock' in~more general cases, which involve more matter components coupled to~each other and~to geometry.

In~the~present paper the~dynamical collapse of~an~uncharged complex scalar field within the~Brans-Dicke theory was considered in~the~context of~time quantification using dynamical scalar fields. The~assumed model also enabled us to~analyze the~results of~a~gravitational evolution of~a~self-interacting real scalar field in~the~Brans-Dicke theory. The~structures of~dynamical spacetimes which form during the~investigated process were described. The~main objective of~the~research was to~examine whether the~scalar field and~the Brans-Dicke field can be used as~time variables in~the vicinity of~the~emerging singularity, which is a~region of~high curvature, as~such regions are of~crucial importance for the~quantum gravity applications. The~conducted studies focused on~the~role of~couplings among the~components of~the~considered model in~order to~assess their significance when scalar fields are to~be used as~time measurers in~the~dynamical system. The~gravitational collapse itself is extensively studied in~quantum gravity~\cite{Torres2014-169,Torres2015-245,Vaz2015-558,GambiniPullin2015-19}. From the~viewpoint of~the~field behavior in~a~spacetime, its course is more complicated than the~usually studied spatially homogeneous cosmological evolution.

The~Brans-Dicke theory was chosen for our studies on~measuring time with the~use of~a~scalar field, because, as~was explained above, it~possesses a~scalar field as~an~intrinsic component which describes gravitational interaction in~combination with the~usual tensorial part. Moreover, it~can be obviously supplied by additional scalar fields in~the~matter sector of~the~constructed theory. Until now, the~gravitational evolution of~collisionless matter within the~Brans-Dicke theory was examined in~the~3+1~formalism~\cite{SheelShapiroTeukolsky1995-4208,SheelShapiroTeukolsky1995-4236}. It~was found that the~final stationary state left after the~process resembles the~one achieved in~general relativity, while the~dynamics of~the~spacetime structure differs significantly from the~results obtained within the~Einstein theory. The~gravitational collapse of~a~real scalar field coupled to~the~Brans-Dicke field was studied in~the~2+2~formalism in~the~context of~the~dependence of~the~emerging spacetime structures on~the~model parameters, dynamical and~late-time behaviors of~the~Brans-Dicke field and~the values of~the~stress-energy tensor components in~the~forming spacetimes~\cite{HwangYeom2010-205002}. The~evolution of~a~gravitationally self-interacting electrically charged scalar field in~the~Brans-Dicke theory was examined in~\cite{HansenYeom2014-040,HansenYeom2015-019}. The~causal structures and~geometries of~the~emerging spacetimes, the~Brans-Dicke field behavior during and~after the~process, the~stress-energy tensor components in~the~dynamical spacetimes were analyzed and~the mass inflation phenomenon was extensively discussed.

The~structure of~the~current paper is the~following. The~theoretical formulation of the~problem is introduced in~section~\ref{sec:bd-model}. Section~\ref{sec:details} contains essential details of~numerical computations and~the presentation of~results. The~main research outcomes are presented and~discussed in~sections~\ref{sec:bd-structures} and~\ref{sec:bd-fielddyn}, while conclusions are gathered in~section~\ref{sec:conclusions}. Appendix~\ref{sec:appendix} is devoted to~the~particulars of~numerical computations and~the code tests.

\section{Brans-Dicke theory with a~scalar field -- theoretical setup}
\label{sec:bd-model}

\subsection{Evolution equations}
\label{sec:bd-model-eqns}

The~action of~the~Brans-Dicke theory with a~complex scalar field is
\ben\label{eqn:BDaction}
S^{BD} = \int d^4 x \sqrt{-g} \left[ \frac{1}{16\pi} 
\left( \Phi R - \frac{\omega}{\Phi} \Phi\Pm \Phi\Pn g^{\mu\nu} \right) + \Phi^\beta L^{SF} \right],
\een
where $\Phi$ denotes the~Brans-Dicke field, $R$ is the~Ricci scalar and~$g$ is the~determinant of~the~metric $g_{\mu\nu}$. The~Lagrangian density of~the~complex scalar field $\phi$ has the~form
\ben\label{eqn:BDaction-lag}
L^{SF} = -\phi\Pm \bp\Pn g^{\mu\nu}.
\een
There are two coupling constants in~the~theory, namely the~Brans-Dicke coupling $\omega$ and the~constant $\beta$, which characterizes the~coupling between the~Brans-Dicke field and~the~complex scalar field. The~above formulation of~the~theory allows us to~investigate the~evolution of~both real and~complex scalar fields within the~Brans-Dicke theory, which will be elaborated in~detail in~section~\ref{sec:details}. The~computations were conducted assuming $c=1$ in~the~Jordan frame, which is often regarded as~physical when considering the~Brans-Dicke theory~\cite{BransDicke1961-925,BarvinskyKamenshchikStarobinsky2008-021}.

The~Einstein equations resulting from the~action~\eqref{eqn:BDaction} can be written as
\ben\label{eqn:Ein-eqns}
G_{\mu\nu} = 8\pi \left( T_{\mu\nu}^{BD} + \Phi^{\beta-1} T_{\mu\nu}^{SF} \right),
\een
where $G_{\mu\nu}$ is the~Einstein tensor and~the~stress-energy tensor components related to~the~Brans-Dicke and~scalar fields, respectively, are
\ben\label{eqn:set-BDfield}
T_{\mu\nu}^{BD} &=& \frac{1}{8\pi\Phi} \left( \Phi_{;\mu\nu} - g_{\mu\nu}\Phi_{;\rho\sigma} g^{\rho\sigma} \right)
+ \frac{\omega}{8\pi\Phi^2} \left( \Phi\Pm \Phi\Pn - \frac{1}{2} g_{\mu\nu} \Phi_{;\rho} \Phi_{;\sigma} g^{\rho\sigma} \right), \\
\label{eqn:set-field}
T_{\mu\nu}^{SF} &=& \phi\Pm \bp\Pn + \bp\Pm \phi\Pn + g_{\mu\nu} L^{SF}.
\een

The~equations of~motion of~the~Brans-Dicke and~scalar fields derived from the~variational principle are
\ben\label{eqn:BDfield-eqn}
\Phi_{;\mu\nu} g^{\mu\nu} - \frac{8\pi\Phi^\beta}{3+2\omega} \left( T^{SF} - 2\beta L^{SF} \right) = 0, \\
\label{eqn:field-eqn}
\phi_{;\mu\nu} g^{\mu\nu} + \frac{\beta}{\Phi} \Phi\Pm \phi\Pn g^{\mu\nu} = 0,
\een
where $T^{SF}$ is the~trace of~\eqref{eqn:set-field}.

\subsection{Double null formalism implementation}
\label{sec:bd-model-dnformalism}

The~spherically symmetric dynamical evolution was traced in~double null coordinates ($u$,~$v$, $\theta$,~$\varphi$), in~which the~general line element has the~form
\ben
ds^2 = -\alpha\left(u,v\right)^2 dudv +r\left(u,v\right)^2 d\Omega^2,
\een
where~$u$ and~$v$ are retarded and~advanced times, respectively, $d\Omega^2=d\theta^2+\sin^2\theta d\varphi^2$ is the~line element of~a~unit sphere, while $\theta$ and~$\varphi$ denote angular coordinates. The~quantities $\alpha$ and~$r$ are arbitrary functions depending on~both the~retarded and~advanced time which reflect the~dynamical evolution of~spacetime in~the~investigated matter-geometry system.

For~convenience of~numerical computations, a~set of~variables was introduced at~the~stage of~deriving the~equations of~motion governing the~evolution of~the~dynamical fields
\ben\label{eqn:subst}
\begin{split}
h &=\frac{\alpha\Pu}{\alpha}, \quad d=\frac{\alpha\Pv}{\alpha}, \quad f=r\Pu, \quad g=r\Pv,\\
W &=\Phi\Pu, \quad Z=\Phi\Pv, \quad w=s\Pu, \quad z=s\Pv,
\end{split}
\een
where $s\equiv\sqrt{4\pi}\phi$ is the~rescaled complex scalar field function.

The~elements of~the~Einstein tensor in~double null coordinates are
\ben
G_{uu} &=& -\frac{2}{r} \left( f\Pu - 2fh \right), \\
G_{vv} &=& -\frac{2}{r} \left( g\Pv - 2gd \right), \\
G_{uv} &=& \frac{1}{2r^2} \left( 4rf\Pv + \alpha^2 + 4fg \right), \\
G_{\theta\theta} &=& \sin^{-2}\theta\ G_{\varphi\varphi} = -\frac{4r^2}{\alpha^2} \left( d\Pu + \frac{f\Pv}{r} \right),
\een
while the~non-zero elements of~the~stress-energy tensor components~\eqref{eqn:set-BDfield} and~\eqref{eqn:set-field} are
\ben
T_{uu}^{BD} &=& \frac{1}{8\pi\Phi} \left( W\Pu - 2hW \right) + \frac{\omega}{8\pi\Phi^2}W^2, \\
T_{vv}^{BD} &=& \frac{1}{8\pi\Phi} \left( Z\Pv - 2dZ \right) + \frac{\omega}{8\pi\Phi^2}Z^2, \\
T_{uv}^{BD} &=& -\frac{Z\Pu}{8\pi\Phi} - \frac{gW + fZ}{4\pi r\Phi}, \\
T_{\theta\theta}^{BD} &=& \sin^{-2}\theta\ T_{\varphi\varphi}^{BD} = \frac{r^2}{2\pi\alpha^2 \Phi}Z\Pu
+ \frac{r}{4\pi\alpha^2 \Phi} \left( gW + fZ \right) + \frac{\omega r^2}{4\pi\Phi^2\alpha^2}WZ, \\
T_{uu}^{SF} &=& \frac{1}{2\pi} w\bw, \\
T_{vv}^{SF} &=& \frac{1}{2\pi} z\bz, \\
T_{\theta\theta}^{SF} &=& \sin^{-2}\theta\ T_{\varphi\varphi}^{SF} = \frac{r^2}{2\pi\alpha^2}
\left( w\bz + z\bw \right).
\een

The~$\theta\theta$ (or $\varphi\varphi$) and~$uv$ components of~the~Einstein equations~\eqref{eqn:Ein-eqns}, together with the~equation of~motion of~the~Brans-Dicke field~\eqref{eqn:BDfield-eqn}, can be written collectively in~a~matrix form
\ben\label{eqn:matrix-1}
\begin{pmatrix}
1\ & \frac{1}{r}\ & \frac{1}{\Phi} \\
0\ & 1\ & \frac{r}{2\Phi} \\
0\ & 0\ & r
\end{pmatrix}
\begin{pmatrix}
d\Pu \\
f\Pv \\
Z\Pu
\end{pmatrix}
=
\begin{pmatrix}
\mathcal{A} \\
\mathcal{B} \\
\mathcal{C}
\end{pmatrix}.
\een
The~elements of~the~right-hand side vector are defined as
\ben
\mathcal{A} &\equiv& -\frac{2\pi\alpha^2}{r^2\Phi} \widetilde{T}_{\theta\theta}^{SF} - \frac{1}{2r\Phi} \left( gW+fZ \right)
- \frac{\omega}{2\Phi^2}WZ, \\
\mathcal{B} &\equiv& -\frac{\alpha^2}{4r} - \frac{fg}{r} + \frac{4\pi r}{\Phi} \widetilde{T}_{uv}^{SF}
- \frac{1}{\Phi} \left( gW+fZ \right), \\
\mathcal{C} &\equiv& -fZ -gW - \frac{2\pi r\alpha^2}{3+2\omega} \left( \widetilde{T}^{SF} - 2\beta\widetilde{L}^{SF} \right),
\een
where $\widetilde{Q}\equiv\Phi^\beta Q$ for any quantity $Q$ and
\ben
T^{SF} &=& -\frac{4}{\alpha^2} T_{uv}^{SF} + \frac{2}{r^2} T_{\theta\theta}^{SF}, \\
L^{SF} &=& \frac{1}{2\pi\alpha^2} \left( w\bz+z\bw \right).
\een
An~equivalent form of~\eqref{eqn:matrix-1} suitable for numerical computations is
\ben\label{eqn:matrix-2}
\begin{pmatrix}
d\Pu = h\Pv \\
r_{,uv} \\
\Phi_{,uv}
\end{pmatrix}
= \frac{1}{r^2}
\begin{pmatrix}
r^2\ & -r\ & -\frac{r}{2\Phi} \\
0\ & r^2\ & -\frac{r^2}{2\Phi} \\
0\ & 0\ & r
\end{pmatrix}
\begin{pmatrix}
\mathcal{A} \\
\mathcal{B} \\
\mathcal{C}
\end{pmatrix}.
\een

The~$uu$ and~$vv$ components of~the~Einstein equations~\eqref{eqn:Ein-eqns} yield the~constraint equations
\ben
r_{,uu} &=& 2fh - \frac{r}{2\Phi} \left( W\Pu - 2hW \right) - \frac{r\omega}{2\Phi^2}W^2
- \frac{4\pi r}{\Phi}\widetilde{T}_{uu}^{SF}, \\
r_{,vv} &=& 2dg - \frac{r}{2\Phi} \left( Z\Pv - 2dZ \right) - \frac{r\omega}{2\Phi^2}Z^2
- \frac{4\pi r}{\Phi}\widetilde{T}_{vv}^{SF},
\een
and the~evolution equation of~the~scalar field~\eqref{eqn:field-eqn} is
\ben\label{eqn:field-eqn-uv}
s_{,uv} &=& -\frac{fz}{r} - \frac{gw}{r} - \frac{\beta}{2\Phi} \left( Wz + Zw \right).
\een

\section{Details of~computer simulations and~results analysis}
\label{sec:details}

\begin{figure}[tbp]
\centering
\includegraphics[width=0.3\textwidth]{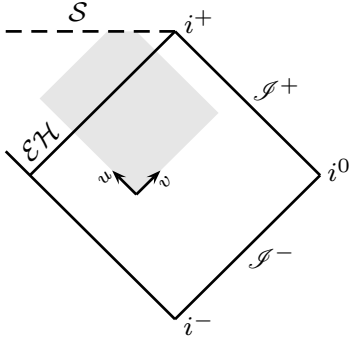}
\caption{The~computational domain (marked gray) on~the~background of~the~Carter-Penrose diagram of~the~Schwarzschild spacetime. The~central spacelike singularity along $r=0$ and~the~event horizon are denoted as~$\mathcal{S}$ and~$\mathcal{EH}$, respectively. $\mathscr{I}^\pm$ and~$i^\pm$ are null and~timelike infinities, while $i^0$ is~a~spacelike infinity.}
\label{fig:domain}
\end{figure}

The~equations~\eqref{eqn:matrix-2}--\eqref{eqn:field-eqn-uv} govern the~dynamical evolution of~the~investigated matter-geometry system, which is a~complex scalar field in~the~Brans-Dicke theory. The~equations were solved numerically within the~spacetime domain situated in~the~$(vu)$-plane, shown in~figure~\ref{fig:domain}. The~exact ranges of~the~null coordinates adopted for numerical calculations were $v\in [0,60]$ and~$u\in [0,20]$. The~code used during~the~simulations and~its consistency checks for the~currently studied case are presented in~appendix~\ref{sec:appendix}.

\subsection{Initial data and~evolution parameters}
\label{sec:iniparam}

The~only arbitrary initial conditions of~the~process were the~field profiles posed on~the~initial $u=const.$ hypersurface. The~profile of~the~evolving complex field was
\ben\label{eqn:field-ini}
\phi\left(v\right) = \frac{A}{\sqrt{8\pi}} \sin^4 \left(\pi\frac{v-v_i}{v_f-v_i}\right) 
\left[\cos\left(2\pi\frac{v-v_i}{v_f-v_i}\right) + i\cos\left(2\pi\frac{v-v_i}{v_f-v_i}-\pi\delta\right)\right]
\een
within the~range $v_i\leqslant v\leqslant v_f$, where $v_i$ and~$v_f$ were equal to~$0$ and~$20$, respectively, and~was equal to~zero outside the~specified range of~advanced time. Due to~the~fact that the~field function is non-zero within the~closed interval of~advanced time, the~spacetime region from within the~specified range will be referred to~as~dynamical. The~parameter $\delta$ takes values from the~range $[0,0.5]$ and~when it~vanishes the~evolving field $\phi$ can be interpreted as~a~real scalar field. The~value of~the~amplitude~$A$ responsible for the~field gravitational self-interaction strength was set as~equal to~$0.25$. The~effective gravitational constant $G=\Phi^{-1}$ was assumed to~be equal to~unity asymptotically and~for this reason the~initial profile of~the~Brans-Dicke field was set as
\ben\label{eqn:BDfield-ini}
\Phi\left(v\right)=1.
\een

The~free evolution parameters $\beta$ and~$\omega$ are model-dependent, while $\delta$, as~was mentioned above, controls the~type of~the~collapsing scalar field. The~considered values of~these constants and~the~corresponding theoretical models are the~following.
\begin{itemize}
 \item $\beta$: $0$ (type IIA~model), $0.5$ (type I model), $1$ (heterotic model),
 \item $\omega$: $10$ (large $\omega$ limit), $0$ ($f(R)$ limit), $-1$ (dilatonic limit), $-1.4$ (brane-world limit), $-1.6$ (ghost limit),
 \item $\delta$: $0$ (real scalar field), $0.5$ (complex scalar field).
\end{itemize}
The~adequate values of~the~free parameters will be given above each diagram presenting the~collapse results.

Specific arguments for the~above classification of~the values of~the constant $\beta$, which originate from a~set of~the string theory-inspired models, and~the~motivations for the~studied values of~$\omega$ are based on~the~classification of~the~full version of~the~string theory, which obviously involves terms of~higher order in~curvature and~contains gauge fields~\cite{Becker}.

The~expansions of~the~effective actions of~the~bosonic sector of~the~type IIA, type I and~heterotic theories are
\ben
S_{\textrm{IIA}} &=& \frac{1}{2\lambda_s^{\, 8}} \int d^{10}x\; \sqrt{-g}\; \Bigg\{ e^{-\phi_d} \left[ R + \big( \nabla\phi_d \big)^2 - \frac{H_3^{\, 2}}{12} \right] - \left( \frac{F_2^{\, 2}}{4} + \frac{\tilde{F}_4^{\, 2}}{48} \right) \Bigg\} 
+ \cdots, \\
S_{\textrm{I}} &=& \frac{1}{2\lambda_s^{\, 8}} \int d^{10}x\; \sqrt{-g}\; \Bigg\{ e^{-\phi_d} \left[ R + \big( \nabla\phi_d \big)^2 \right] - \frac{\tilde{H}_3^{\, 2}}{12} - e^{-\frac{\phi_d}{2}} \frac{\textrm{Tr}\big(F_2^{\, 2}\big)}{4} \Bigg\} 
+ \cdots, \\
S_{\textrm{het}} &=& \frac{1}{2\lambda_s^{\, 8}} \int d^{10}x\; \sqrt{-g}\; e^{-\phi_d} \left[ R + \big( \nabla\phi_d \big)^2 - \frac{\tilde{\tilde{H}}_3^{\, 2}}{2} - \frac{\textrm{Tr}\big(F_2^{\, 2}\big)}{4} \right] + \cdots,
\een
where $\lambda_s$ is the~length scale of~strings and~$\phi_d$ denotes a~dilaton field. $H_3$ is the~field strength tensor of~the~NS-NS two-form $B_2$, while $\tilde{H}_3$ and~$\tilde{\tilde{H}}_3$ stand for mixed contributions of~the~R-R two-form $A_2$ and~the~NS-NS two-form $B_2$, respectively, and~the~matrix-valued one-form~$A_1$. $F_2$ is the~field strength tensor of~the~R-R one-form $A_1$ and~$\tilde{F}_4=dA_3+A_1\land H_3$ with $A_3$ being a~three-form field. The~dimensional reduction procedure results in~the~effective actions for the~considered theories, which can be collectively written in~the~following form:
\ben
S_{\textrm{IIA/I/het}}^{(4)} = \frac{1}{16\pi} \int d^4x\; \sqrt{-g}\; \bigg\{ e^{-\phi_d} \left[ R + \big( \nabla\phi_d \big)^2 \right] - \rchi F_2^{\, 2} \bigg\} + \cdots,
\een
where $\rchi$ equals $1$ for the~type~IIA, $e^{-\frac{\phi_d}{2}}$ for the~type~I and~$e^{-\phi_d}$ for the~heterotic theory. The~redefinition of~the~dilaton field $e^{-\phi_d}\to\Phi_d$ leads to~the~gravitational sector, which is identical for all the~examined versions of~the~string theory and~proportional to~the~term $\Phi_d \Big[ R + \big( \nabla\Phi_d \big)^2 \Phi_d^{\: -2} \Big]$. The~two-form field becomes proportional to~the~term $\Phi_d^{\:\beta} F_2^{\, 2}$ with $\beta$ equal to~$0$ for the~type~IIA, $0.5$ for the~type~I and~$1$ for the~heterotic theory. The~obtained couplings provided an~inspiration for the~values of~the~Brans-Dicke field -- matter sector coupling for the~theoretical setup studied in~the~current paper.

The~large $\omega$ limit was identified with the~Brans-Dicke coupling equal to~$10$, because for its larger values the~behavior of~the~system does not change qualitatively. The~gravitational sector of~the~action which corresponds to~the~scalar-tensor version of~the~$f(R)$ gravity is
\ben
S_{f(R)} = \frac{1}{16\pi} \int d^4x\; \sqrt{-g}\; \Big[ f(\psi) + f^\prime(\psi) \left(R-\psi\right) \Big],
\een
where $\psi$ is an~auxiliary scalar field~\cite{SotiriouFaraoni2010-451}. The~field redefinition $f^\prime(\psi)\to\Phi_\psi$ allows to~write the~above action in~the~form
\ben
S_{f(R)} = \frac{1}{16\pi} \int d^4x\; \sqrt{-g}\; \Big( \Phi_\psi R - V(\Phi_\psi) \Big),
\een
which corresponds to~a~Brans-Dicke field with potential when the~coupling vanishes. Thus, the~case of~$\omega=0$ was interpreted as the~$f(R)$ limit of~the~theory. The~low energy effective action of~each string theory contains a~sector with the~dilaton field
\ben
S_{\textrm{d}} &=& \frac{1}{2\lambda_s^{\, D-1}} \int d^{D+1}x\; \sqrt{-g}\; e^{-\phi_d} \left[ R + \big( \nabla\phi_d \big)^2 \right],
\een
where $D$ denotes the~number of~space dimensions~\cite{Gasperini}. The~field redefinition $\lambda_s^{\, 1-D}\cdot e^{-\phi_d}\to\Phi_d\left(8\pi G_{D-1}\right)^{-1}$, where $G_{D-1}$ is the~$D-1$-dimensional gravitational constant, gives the~$\omega=-1$ limit of~the~Brans-Dicke theory, which was called dilatonic. The~value of~the~Brans-Dicke parameter can be calculated in~the~weak field limit of~the~Randall-Sundrum braneworld model~\cite{RandallSundrum1999-3370} according to
\ben
\omega = \frac{3}{2} \left( e^{\pm\frac{s}{l}}-1 \right),
\een
where $s$ is the~distance between the~branes and~$l=\sqrt{-6\Lambda^{-1}}$ denotes the~length scale of~the~anti-de Sitter space, while the~sign in~the~exponent depends on~the~sign of~the~brane tension~\cite{GarrigaTanaka2000-2778}. The~value of~$\omega$ in~these models is usually close to~$-1.5$. The~value $-1.4$ chosen for our computations thus represents the~brane-world limit. When $\omega$ is less than $-1.5$, the~kinetic term of~the~Brans-Dicke field in~the~Lagrangian is negative in~the~Einstein frame and~hence the~field acts as a~ghost~\cite{KimLeeLeeLeeYeom2011-023519}. The~value of~$-1.6$ was chosen as an~example to~investigate such an~exotic behavior.

\subsection{Penrose and~field diagrams}
\label{sec:diags}

The~outcomes of~the~studied evolutions will be presented through the~prism of~causal structures of~spacetimes and~the~behavior of~fields in~emerging dynamical spacetimes. The~spacetime structures will be depicted on~Penrose diagrams, which present the~$r=const.$ lines in~the~$(vu)$-plane. On all plots of~$r$-contours, the~values of~$r$ range from $0$ to~$40$ and~the~differences between adjacent $r=const.$ lines are $\Delta r=1$. The~central singularity situated along the~$r=0$ line is marked as~a~thick black line. Apparent horizons located along the~hypersurfaces of~a~vanishing expansion $r\Pv=0$ and~$r\Pu=0$ are denoted as~red and~blue curves, respectively. Cauchy horizons, that is null hypersurfaces beyond which the~evolution cannot be extended using the~data from the~previous time slice, are marked as~green lines. The~Penrose diagrams obtained during the~numerical simulations are related via a~conformal transformation to~the~Carter-Penrose diagrams, which depict the~global structure of~spacetimes and~picture the~causal relations within their distinct regions. The~Carter-Penrose diagrams of~the~spacetimes formed in~the~examined process will be also presented.

The~behavior of~fields will be interpreted on~the~basis of~the~field functions values in~the~spacetime. Since one of~the~aims of~the~current studies is to~assess a~possibility of~measuring time intrinsically, that is with the~use of~the fields comprising the~examined dynamical system, the~contours of~constancy hypersurfaces will also be depicted. The~spacetime regions in~which the~constant field function hypersurfaces are spacelike will be marked blue on~the~plots, while the~areas in~which they are timelike will be marked purple. The~values of~field functions covered by the~computations and~the~adequate steps between the~field constancy hypersurfaces are the~following.
\begin{itemize}
 \item For~$\omega > -1.5$, the~Brans-Dicke field ranges from $\Phi = 0.1$ to~$\Phi = 1$ and~the~step $\Delta\Phi$ equals $0.01$.
 \item For~$\omega > -1.5$, the~real part of~the~complex scalar field ranges from $\phi_{Re} = -1$ to~$\phi_{Re} = 0.5$ with the~step $\Delta\phi_{Re} = 0.03$.
 \item For~$\omega < -1.5$, the~range of~the~Brans-Dicke field is $\Phi = 1$ to~$\Phi = 20.9$ and~the~step between adjacent contours is $\Delta\Phi = 0.2$.
 \item For~$\omega < -1.5$, the~real part of~the~complex scalar field ranges from $\phi_{Re} = -2.5$ to~$\phi_{Re} = 0.5$ and~$\Delta\phi_{Re}$ is equal to~$0.02$.
\end{itemize}
The~asymptotic regions and~areas neighboring the~central singularity in~the~dynamical part of~a~spacetime were magnified for several selected diagrams. In~these cases the~field function contours were plotted more densely and~the~values of~steps were placed in~captions of~adequate figures in~section~\ref{sec:bd-fielddyn}.

The~Lagrangian of~the~complex scalar field~\eqref{eqn:BDaction-lag} has a~continuous global symmetry, which rotates the~real and~imaginary parts of~the~field (or,~equivalently, changes the~field phase, $\phi\to e^{i\vartheta}\phi$, where $\vartheta$ is the~phase angle). The~derived equation of~motion~\eqref{eqn:field-eqn} is also invariant, so no physical differences in~the~behavior of~the~real and~imaginary parts of~the~complex scalar field are expected during the~evolution. For~this reason only the~behavior of~the~real field component was shown on~the~plots.

\section{Structures of~spacetimes}
\label{sec:bd-structures}

\begin{figure}[tbp]
\centering
\includegraphics[width=0.8\textwidth]{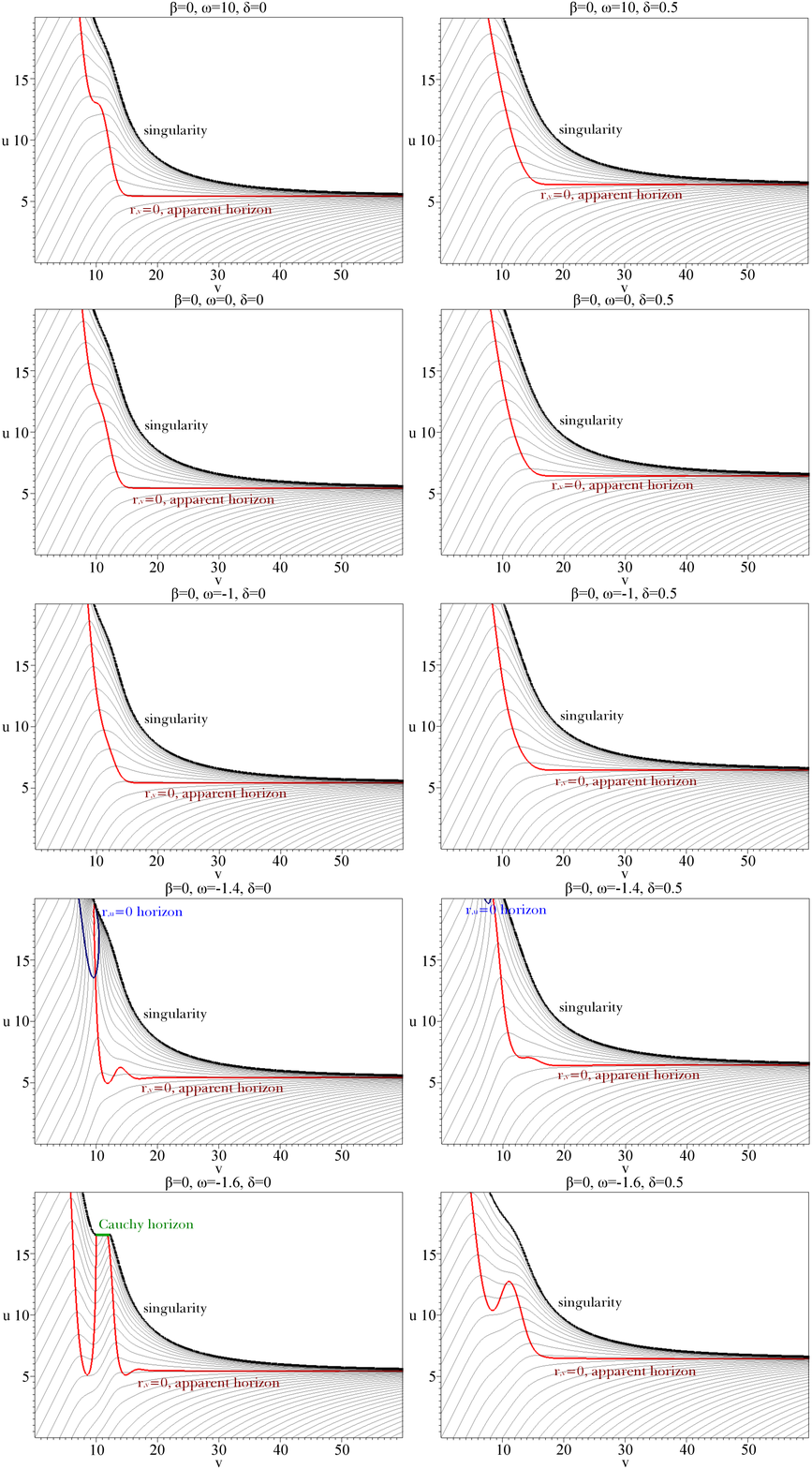}
\caption{(color online) The~Penrose diagrams of~spacetimes formed during evolutions of~real and~complex scalar fields in~the~regime of~the~Brans-Dicke theory for $\beta=0$.}
\label{fig:beta0-r}
\end{figure}

\begin{figure}[tbp]
\centering
\includegraphics[width=0.8\textwidth]{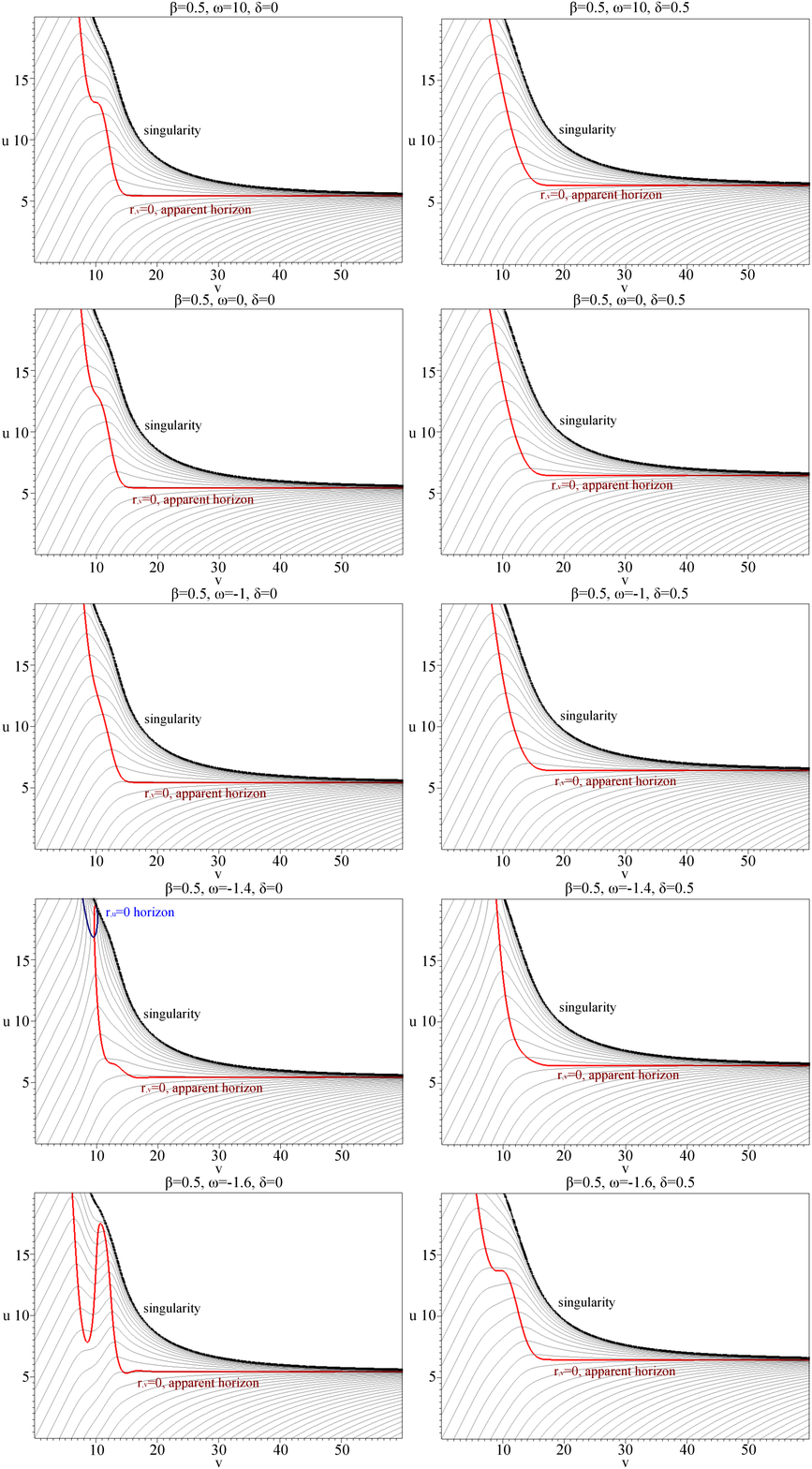}
\caption{(color online) The~Penrose diagrams of~spacetimes formed during evolutions of~real and~complex scalar fields in~the~regime of~the~Brans-Dicke theory for $\beta=0.5$.}
\label{fig:beta05-r}
\end{figure}

\begin{figure}[tbp]
\centering
\includegraphics[width=0.8\textwidth]{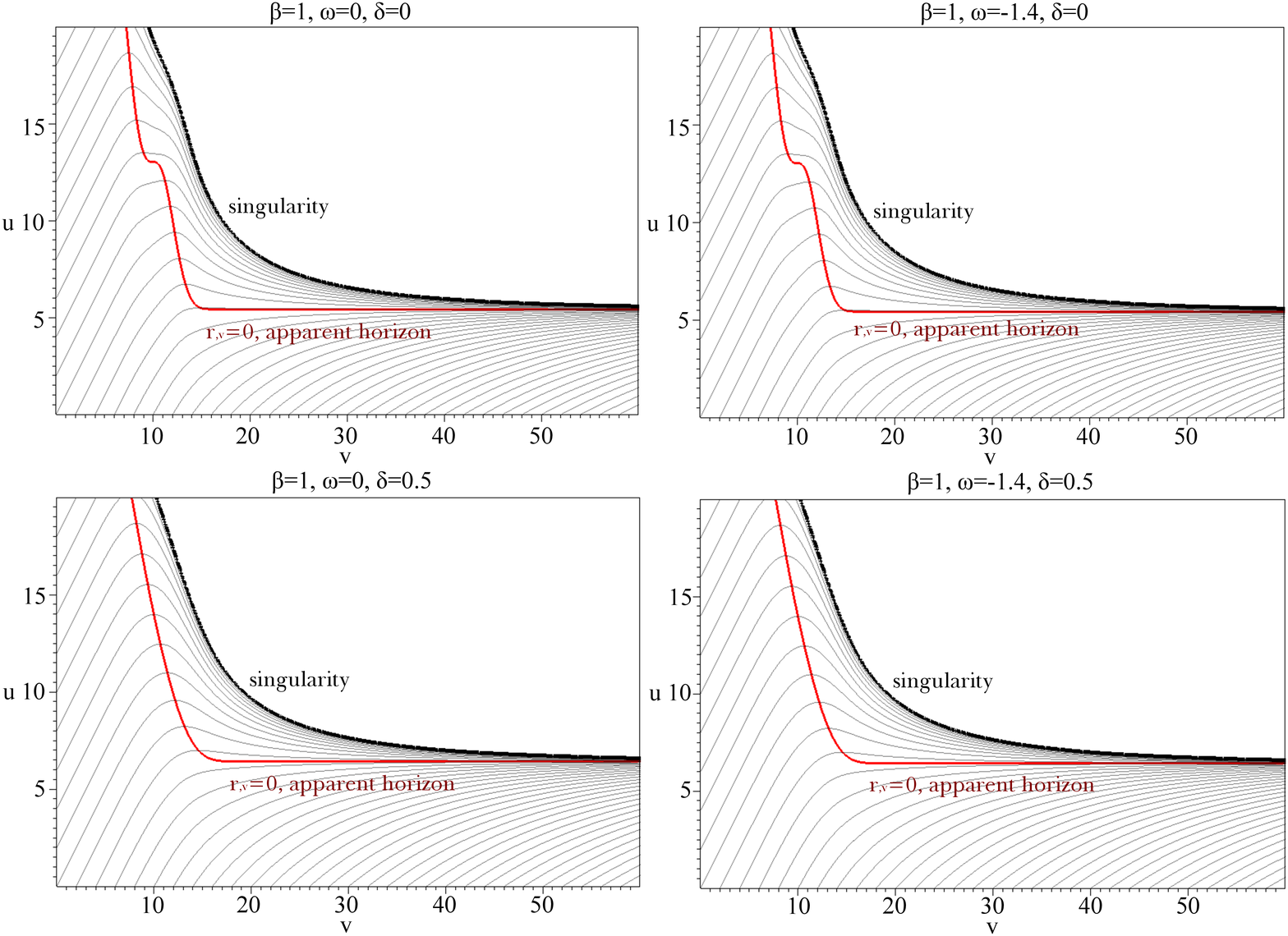}
\caption{(color online) The~Penrose diagrams of~spacetimes formed during evolutions of~real and~complex scalar fields in~the~regime of~the~Brans-Dicke theory for $\beta=1$.}
\label{fig:beta1-r}
\end{figure}

The~structures of~spacetimes resulting from the~gravitational collapse of~real and~complex scalar fields in~the~Brans-Dicke theory are presented in~figures~\ref{fig:beta0-r},~\ref{fig:beta05-r} and~\ref{fig:beta1-r} for the~coupling constant $\beta$ equal to~$0$, $0.5$ and~$1$, respectively.

\subsection{Uncoupled Brans-Dicke and~scalar fields}
\label{sec:struct-uncoupled}

In~the~uncoupled case, i.e.,~for $\beta=0$, the~spacetimes formed during the~evolution of~both real and~complex scalar fields for $\omega$ equal to~$10$, $0$ and~$-1$ contain a~spacelike singularity along $r=0$, which is surrounded by a~single apparent horizon $r\Pv=0$. The~horizon hypersurface is spacelike at~early advanced times and~becomes null as~$v\to\infty$ where the~collapse dynamics settles at~a~final stationary state. The~described structure is also a~result of~the~gravitational collapse of~a~neutral scalar field proceeding in~Einstein gravity~\cite{HamadeStewart1996-497}, the~collapse of~an~electrically charged scalar field in~dilaton gravity~\cite{BorkowskaRogatkoModerski2011-084007,NakoniecznaRogatko2012-3175} and~the gravitational evolution of~a~complex scalar field coupled with phantom Maxwell field~\cite{NakoniecznaRogatkoModerski2012-044043} or~with a~non-phantom Maxwell field and~a~complex scalar field with a~quartic potential charged under a~U(1) gauge field~\cite{NakoniecznaRogatkoNakonieczny2015-012} within general relativity. For~$\omega=-1.4$ the~spacetime which stems from the~investigated evolution contains a~central spacelike singularity encompassed by an~apparent horizon $r\Pv=0$, which can be either spacelike or~timelike in~the~dynamical region of~the~spacetime and~is located along a~null hypersurface $u=const.$ for large advanced times. There also exists another horizon in~the~spacetime, which is situated at~the~$r\Pu=0$ hypersurface and~forms at~late retarded times in~the~dynamical region of~the~spacetime. In~the~ghost limit, for $\omega$ equal to~$-1.6$, the~emerging structures differ for real and~complex scalar fields. In~the~case of~$\delta=0$, there are two portions of~a~central spacelike singularity. They are separately surrounded by the~$r\Pv=0$ apparent horizons, which are spacelike or~timelike in~the~spacetime area, in~which the~collapse dynamics takes place. As~in~all previous cases, the~apparent horizon changes its character to~null as~$v\to\infty$. The~two parts of~the~singularity mentioned above are linked by a~Cauchy horizon null segment, which is not surrounded by any apparent horizon and~thus can be visible for a~distant observer. This violation of~the~weak cosmic censorship conjecture is due to~the~violation of~the~energy conditions in~the~ghost limit. When the~parameter $\delta$ equals $0.5$, the~spacetime contains a~spacelike singularity along $r=0$. It~is surrounded by a~single apparent horizon which is either spacelike or~timelike in~the~region related to~the~field dynamics and~settles along $u=const.$ hypersurface at~large advanced times. The~results obtained for the~real scalar field in~the~type IIA~inspired model are in~agreement with previous findings presented in~\cite{HwangYeom2010-205002}.

\subsection{Coupled Brans-Dicke and~scalar fields}
\label{sec:struct-coupled}

The~course and~outcomes of~the~studied process in~the~type I inspired model, that is for $\beta=0.5$, are qualitatively similar to~the~above-mentioned case of~the~vanishing coupling between the~Brans-Dicke field and~the~scalar field. The~only difference can be noted when $\omega$ equals $-1.4$ and~$-1.6$, because in~contrast to~the~$\beta=0$ case, the~apparent horizon $r\Pu=0$ and~the~Cauchy horizon do not form in~the~spacetime at~late retarded time when $\beta$ takes the~value of~$0.5$ for~$\delta$ equal to~$0.5$ and $0$, respectively.

When $\beta$ is equal to~$1$, the~spacetimes formed during the~examined process for all values of~the~Brans-Dicke coupling constant $\omega$ have similar structures. The~intrinsic spacelike singularity at~$r=0$ is surrounded by an~apparent horizon $r\Pv=0$, which is spacelike for small advanced times and~becomes null as~$v\to\infty$. There exists one dissimilarity between the~evolutions of~real and~complex scalar fields. In~the~case of~vanishing $\delta$, the~spacelike part of~the~horizon passes directly to~the~null part, while when $\delta=0.5$ there is an~intermediate stage, in~which the~horizon is timelike for medium values of~advanced time.

\subsection{Overall dependence on~the~evolution parameters}
\label{sec:struct-sum}

\begin{figure}[tbp]
\subfigure[][]{\includegraphics[width=0.15\textwidth]{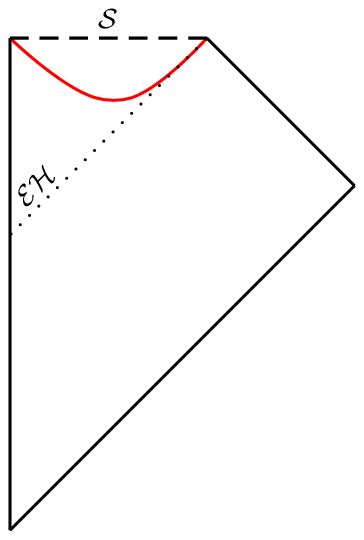}}
\hfill
\subfigure[][]{\includegraphics[width=0.15\textwidth]{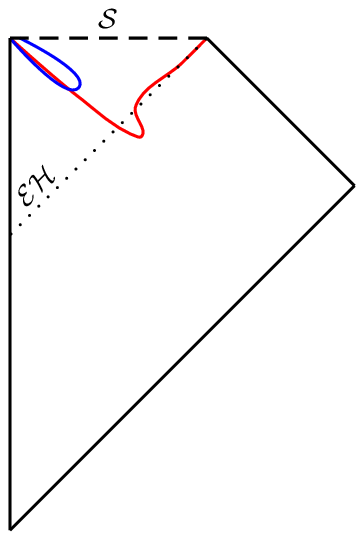}}
\hfill
\subfigure[][]{\includegraphics[width=0.15\textwidth]{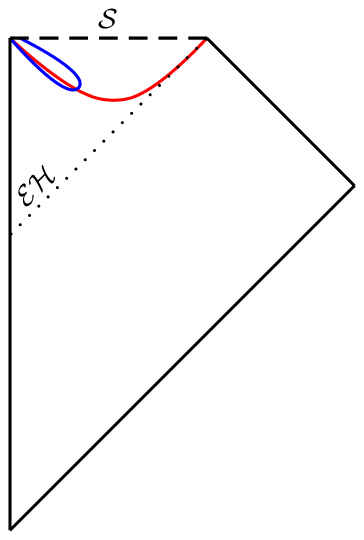}}
\hfill
\subfigure[][]{\includegraphics[width=0.15\textwidth]{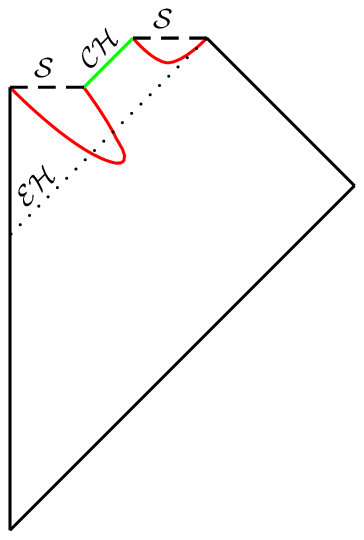}}
\hfill
\subfigure[][]{\includegraphics[width=0.15\textwidth]{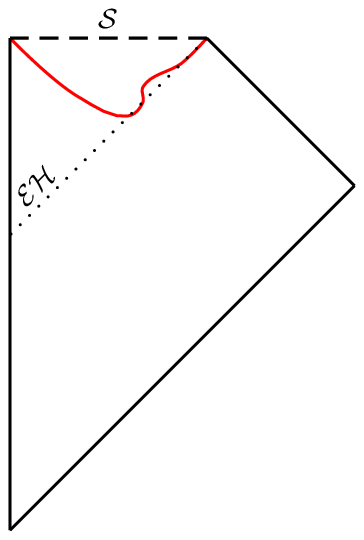}}
\hfill
\subfigure[][]{\includegraphics[width=0.15\textwidth]{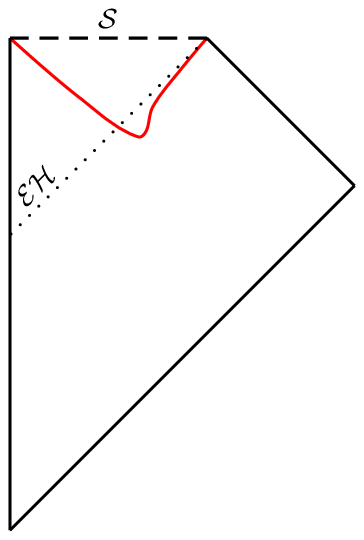}}
\caption{(color online) The~Carter-Penrose diagrams of~spacetimes formed during evolutions of~a~scalar field in~the~Brans-Dicke theory, for the~following sets of~evolution parameters: (a)~$\omega\geqslant -1$ for $\beta=0$, $0.5$ and~$\beta=1$, $\delta=0$, (b)~$\beta=0$, $0.5$, $\omega=-1.4$, $\delta=0$, (c)~$\beta=0$, $\omega=-1.4$, $\delta=0.5$, (d)~$\beta=0$, $\omega=-1.6$, $\delta=0$, (e)~$\beta=0$, $\omega=-1.6$, $\delta=0.5$ and~$\beta=0.5$, $\omega=-1.6$ and~(f)~$\beta=0.5$, $\omega=-1.4$, $\delta=0.5$ and~$\beta=1$, $\delta=0.5$. The~central singularity along $r=0$, the~event and~Cauchy horizons are denoted as~$\mathcal{S}$, $\mathcal{EH}$ and~$\mathcal{CH}$, respectively.}
\label{fig:CPdiags}
\end{figure}

In~general, the~variety and~complexity of~the~spacetime structures formed during the~gravitational evolution of~a~scalar field in~the~Brans-Dicke theory decrease as~the~value of~the~coupling constant $\beta$ between the~Brans-Dicke field and~the~scalar field increases. Moreover, the~model dependence reflected in~the~values of~the~Brans-Dicke coupling $\omega$ indicates that the~collapse of~a~scalar field in~the~large $\omega$, $f(R)$ and~dilatonic limits proceeds similarly to~the~same process in~the~Einstein gravity. In~the~brane-world limit additional apparent horizons are possible at~late times, while the~ghost limit enables the~formation of~more exotic structures, such as~these in~which the~weak cosmic censorship can be violated. A~summary of~causal structures of~spacetimes obtained as~a~result of~the~examined collapse is presented in~figure~\ref{fig:CPdiags} in~the~form of~Carter-Penrose diagrams.

\section{Dynamical behavior of~fields}
\label{sec:bd-fielddyn}

The~evolution of~the~Brans-Dicke and~complex scalar fields~\eqref{eqn:BDfield-eqn}--\eqref{eqn:field-eqn} in~double null coordinates is governed by the~following equations:
\ben\label{eqn:BDfield-eqn2}
\Phi\Puv &=& -\frac{fZ+gW}{r} - \frac{\Phi^\beta\left(1-\beta\right)}{3+2\omega} \left(w\bz+z\bw\right), \\
\label{eqn:field-eqn2}
\phi\Puv &=& -\frac{fz+gw}{r} - \frac{\beta}{2\Phi} \left(Wz+Zw\right).
\een
As~may be inferred from~\eqref{eqn:BDfield-eqn2}, the~case of~$\omega=-1.5$ is a~singular point of~the~evolution equation. The~dynamical behavior of~the~Brans-Dicke field when it~approaches the~central singularity depends on~the~value of~the~parameter $\omega$. For~$\omega > -1.5$, the~Brans-Dicke field moves toward $\Phi = 0$ nearby the~singularity, i.e.,~it~is biased toward the~strong coupling limit. Conversely, for $\omega < -1.5$, it~tends to~$\Phi = \infty$ as~it~evolves toward the~singularity, which means that it~is biased toward the~weak coupling limit.

\subsection{Type IIA~and~type I models}
\label{sec:fields-}

\begin{figure}[tbp]
\centering
\includegraphics[width=0.8\textwidth]{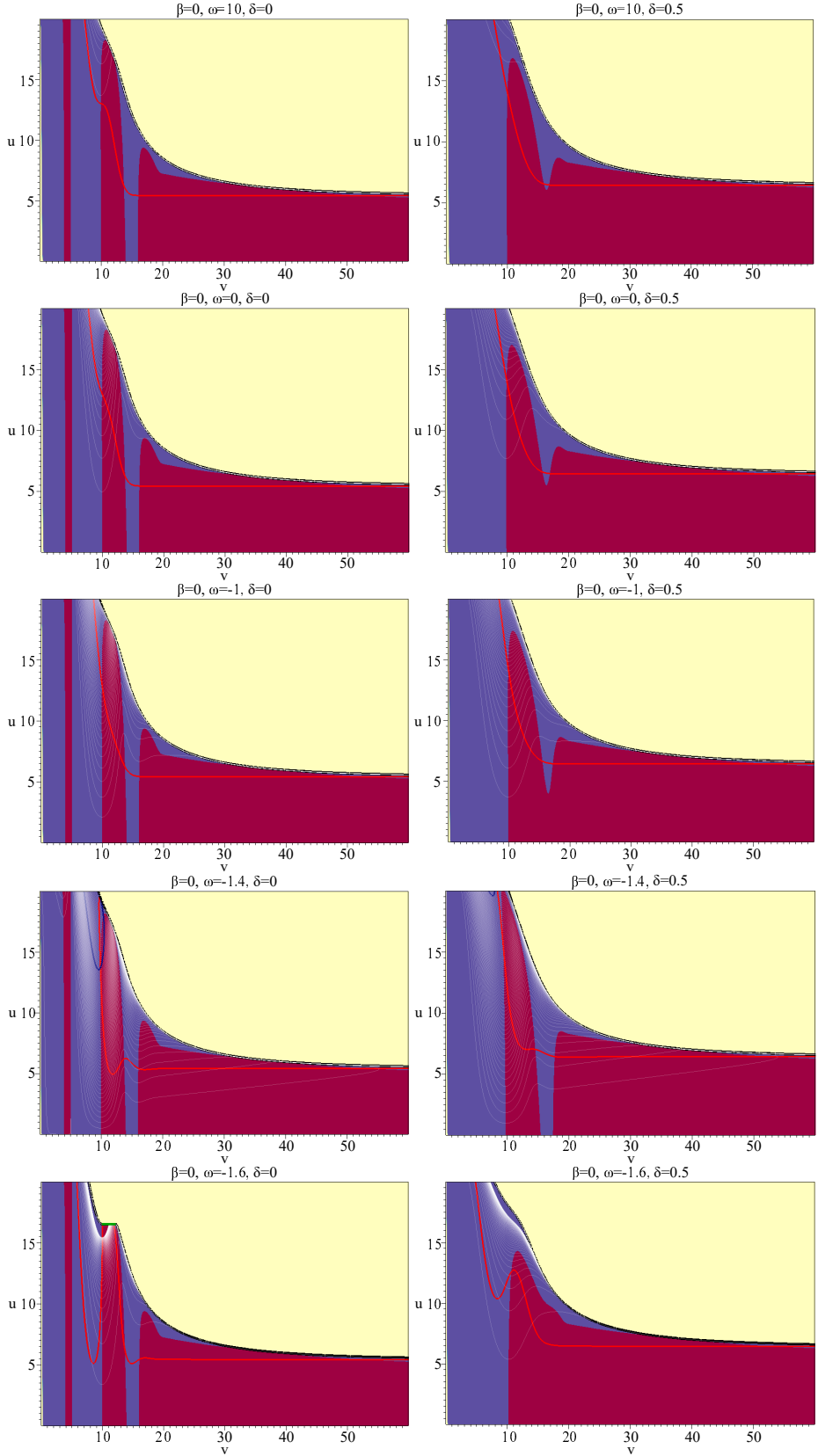}
\caption{(color online) The~constancy hypersurfaces of~the~Brans-Dicke field for evolutions conducted within the~Brans-Dicke theory for $\beta=0$.}
\label{fig:beta0-BD}
\end{figure}

\begin{figure}[tbp]
\centering
\includegraphics[width=0.8\textwidth]{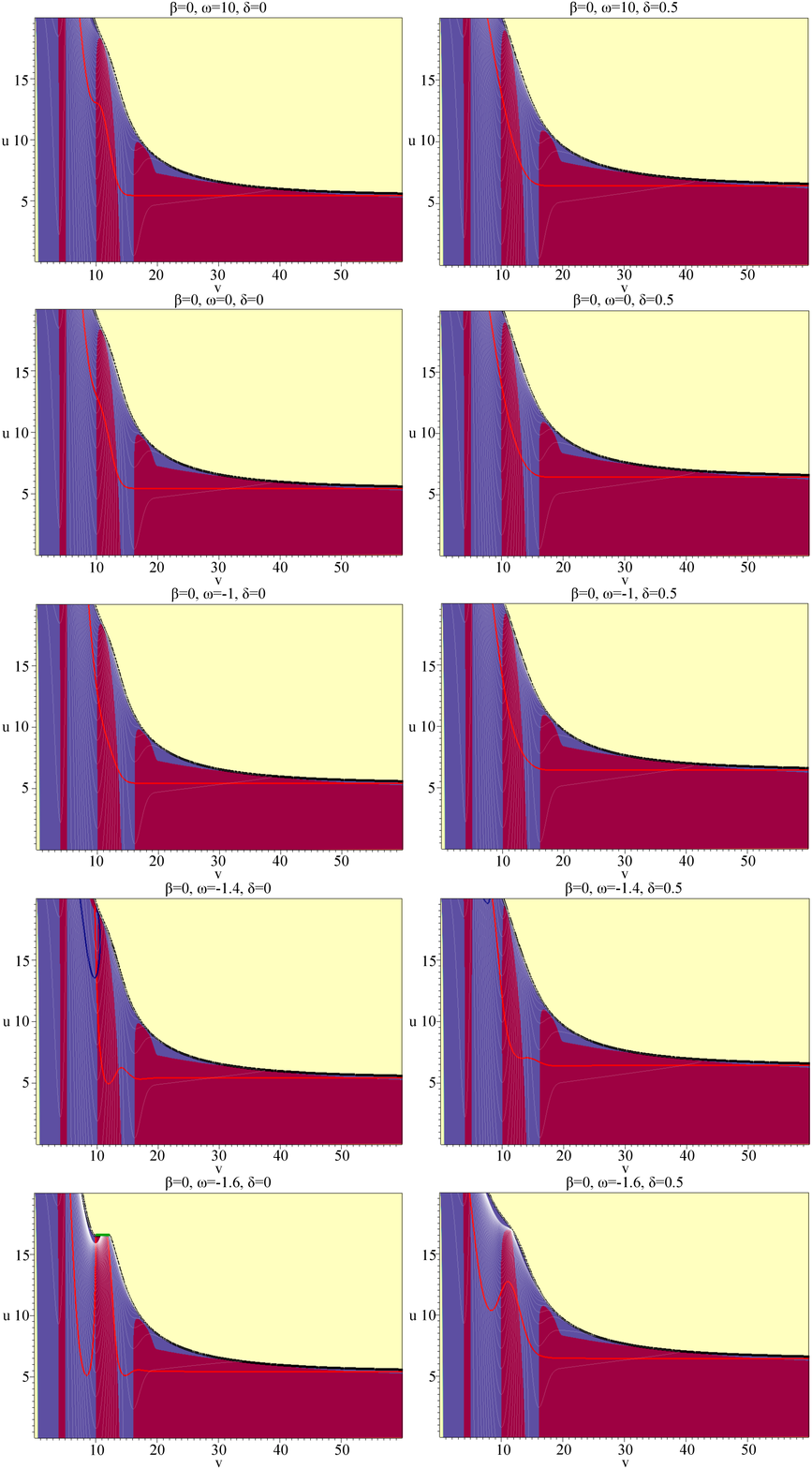}
\caption{(color online) The~contours of~the~real part of~the~complex scalar field for evolutions conducted within the~Brans-Dicke theory for $\beta=0$.}
\label{fig:beta0-re}
\end{figure}

\begin{figure}[tbp]
\subfigure[][]{\includegraphics[width=0.44\textwidth]{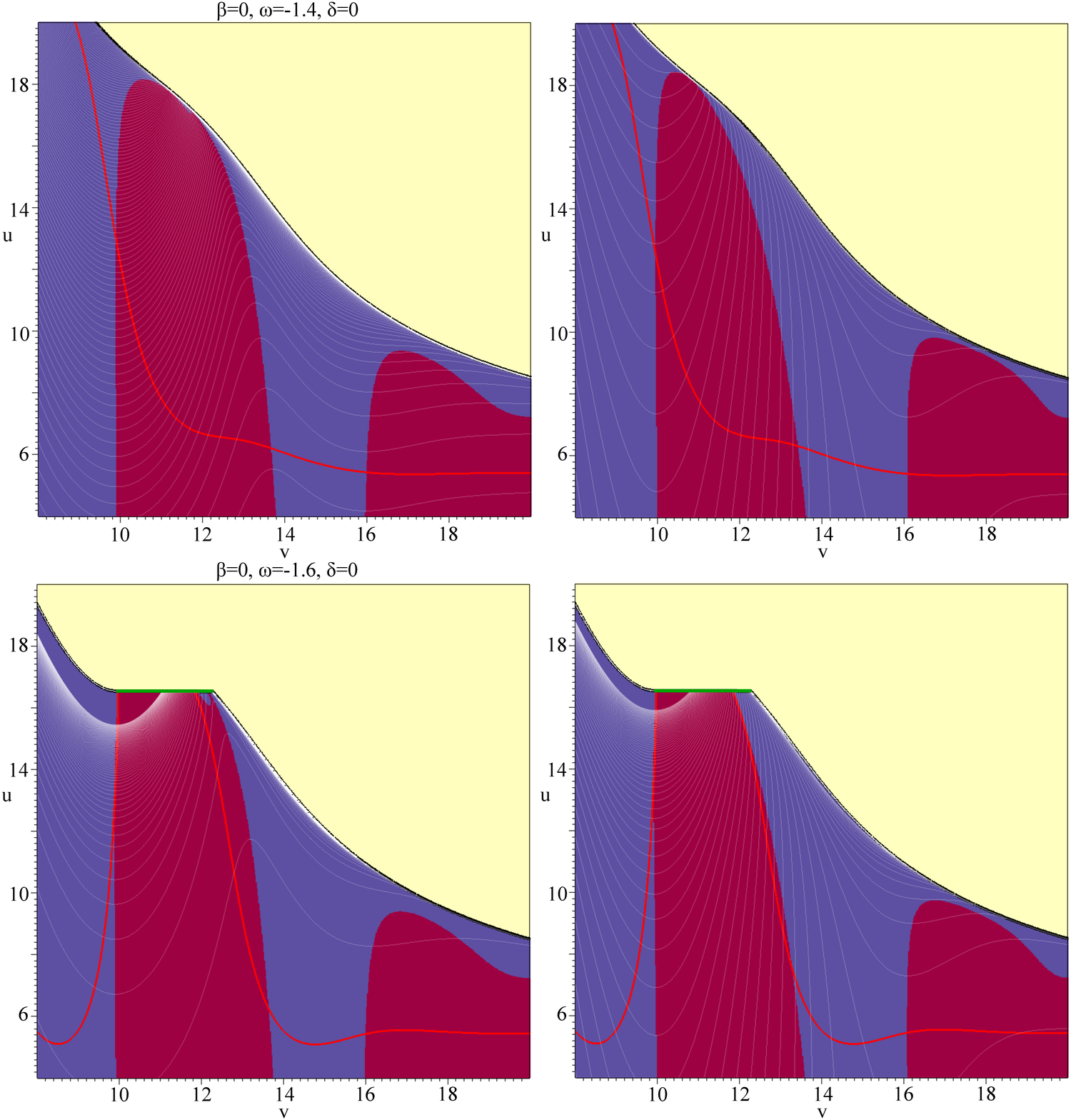}}
\hfill
\subfigure[][]{\includegraphics[width=0.44\textwidth]{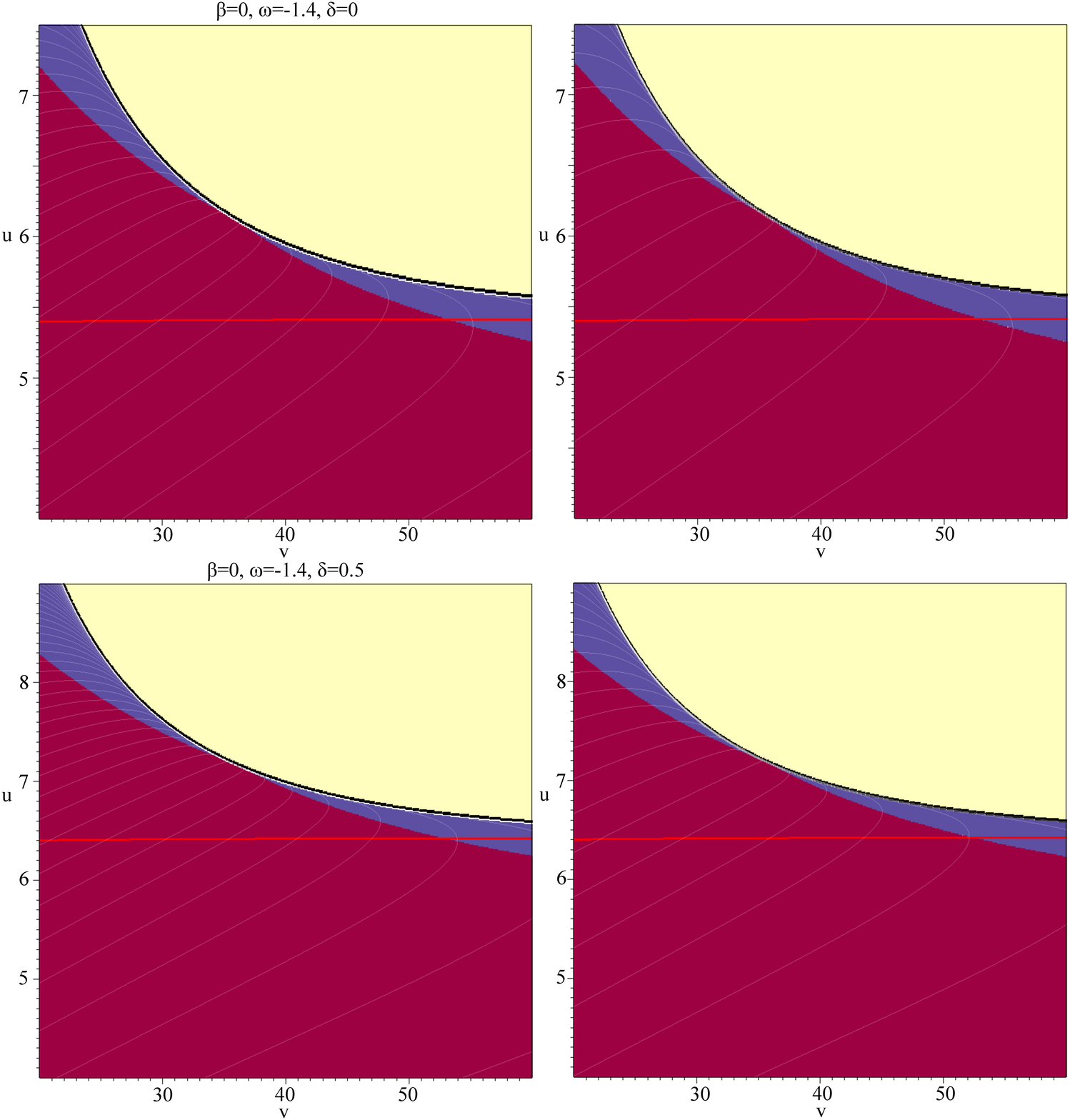}}
\caption{(color online) Field contours in~(a)~the dynamical region and~(b)~the asymptotic region of~spacetimes formed in~the~collapse within the~Brans-Dicke theory for $\beta=0$. Left column: the~Brans-Dicke field. Right column: the~real part of~the~complex scalar field. The~steps between the~contours are $\Delta \Phi = 0.0025$ and~$\Delta \phi_{Re} = 0.002$.}
\label{fig:beta0-enla}
\end{figure}

\begin{figure}[tbp]
\centering
\includegraphics[width=0.8\textwidth]{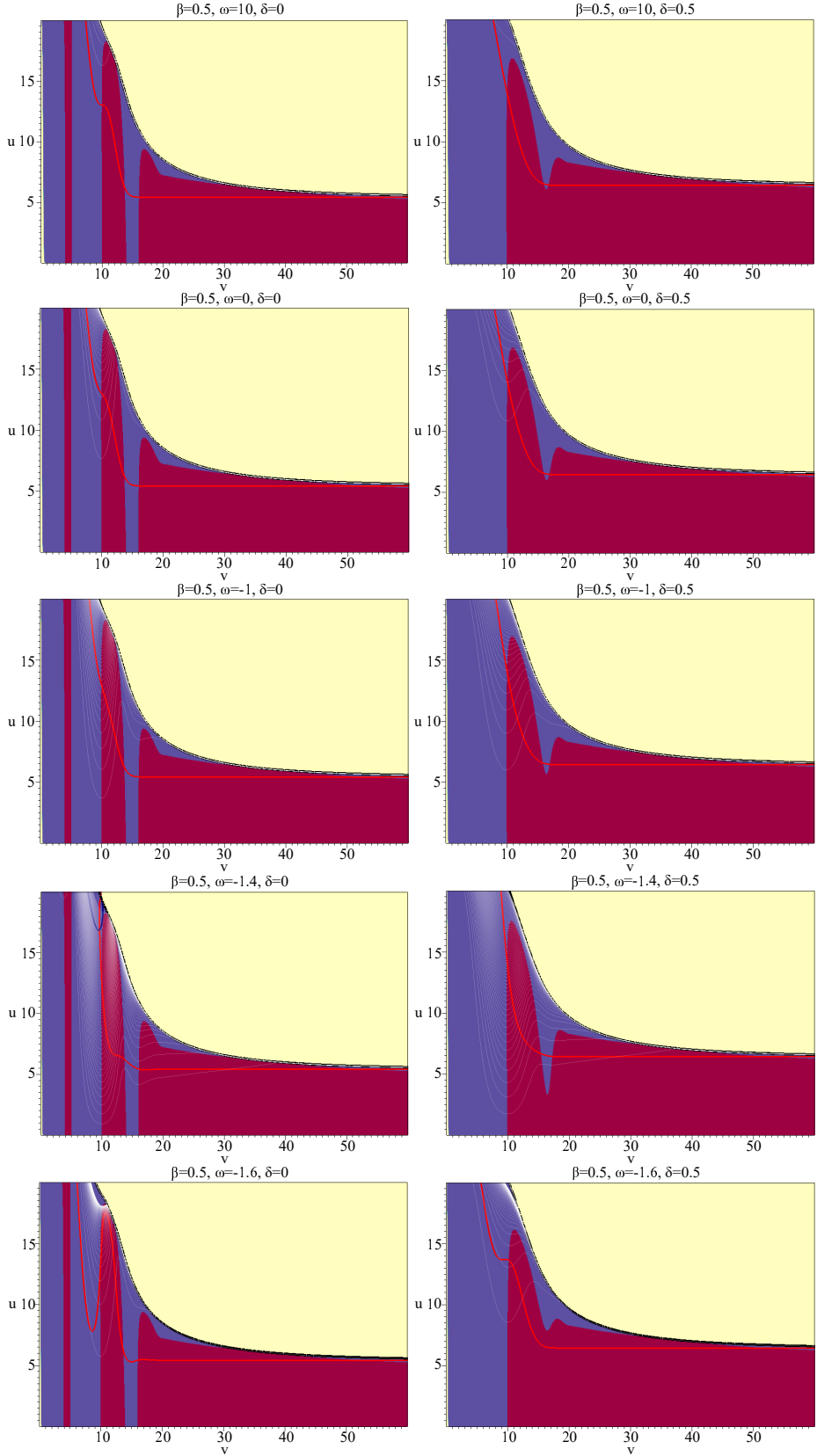}
\caption{(color online) The~constancy hypersurfaces of~the~Brans-Dicke field for evolutions conducted within the~Brans-Dicke theory for $\beta=0.5$.}
\label{fig:beta05-BD}
\end{figure}

\begin{figure}[tbp]
\centering
\includegraphics[width=0.8\textwidth]{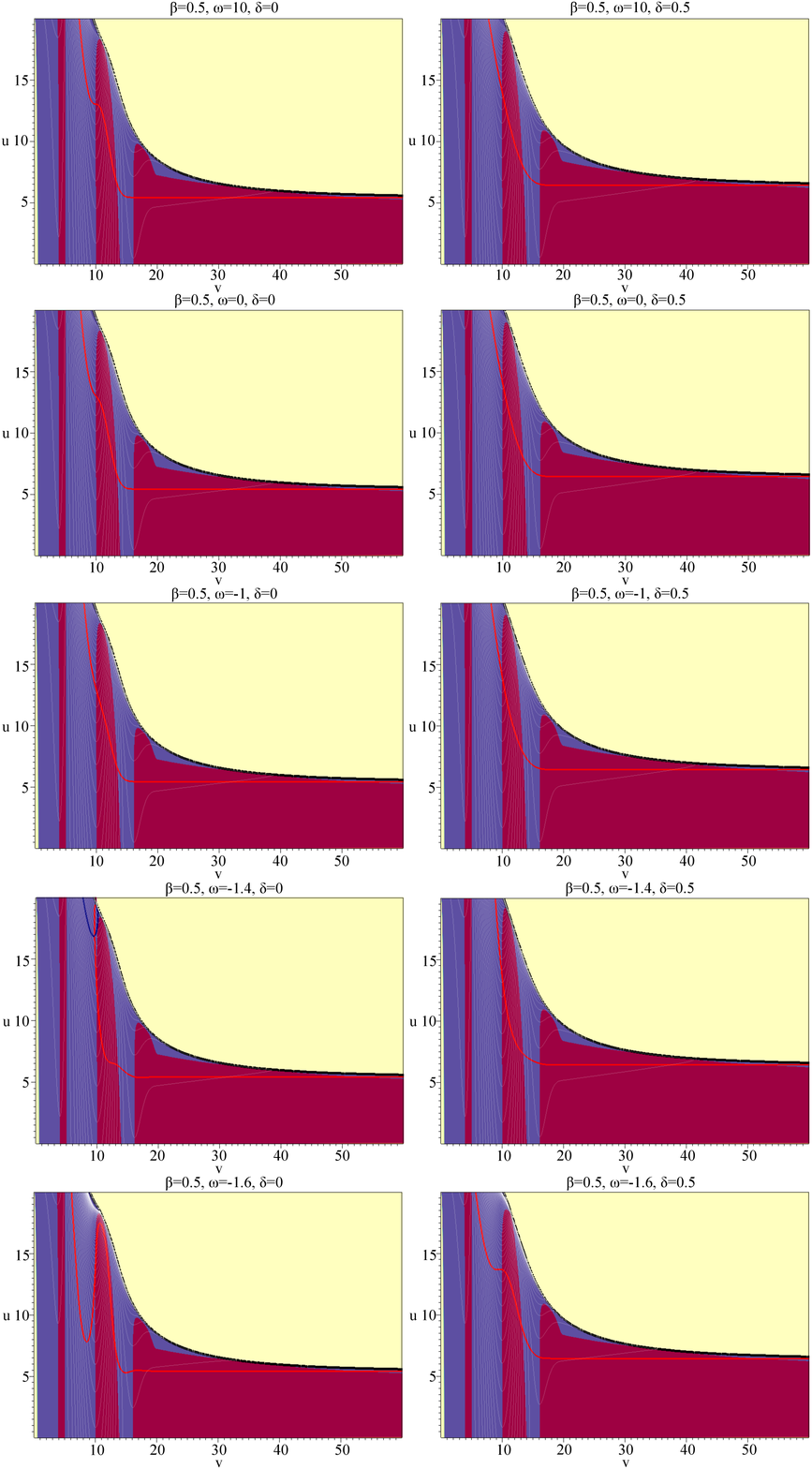}
\caption{(color online) The~contours of~the~real part of~the~complex scalar field for evolutions conducted within the~Brans-Dicke theory for $\beta=0.5$.}
\label{fig:beta05-re}
\end{figure}

\begin{figure}[tbp]
\centering
\includegraphics[width=0.46\textwidth]{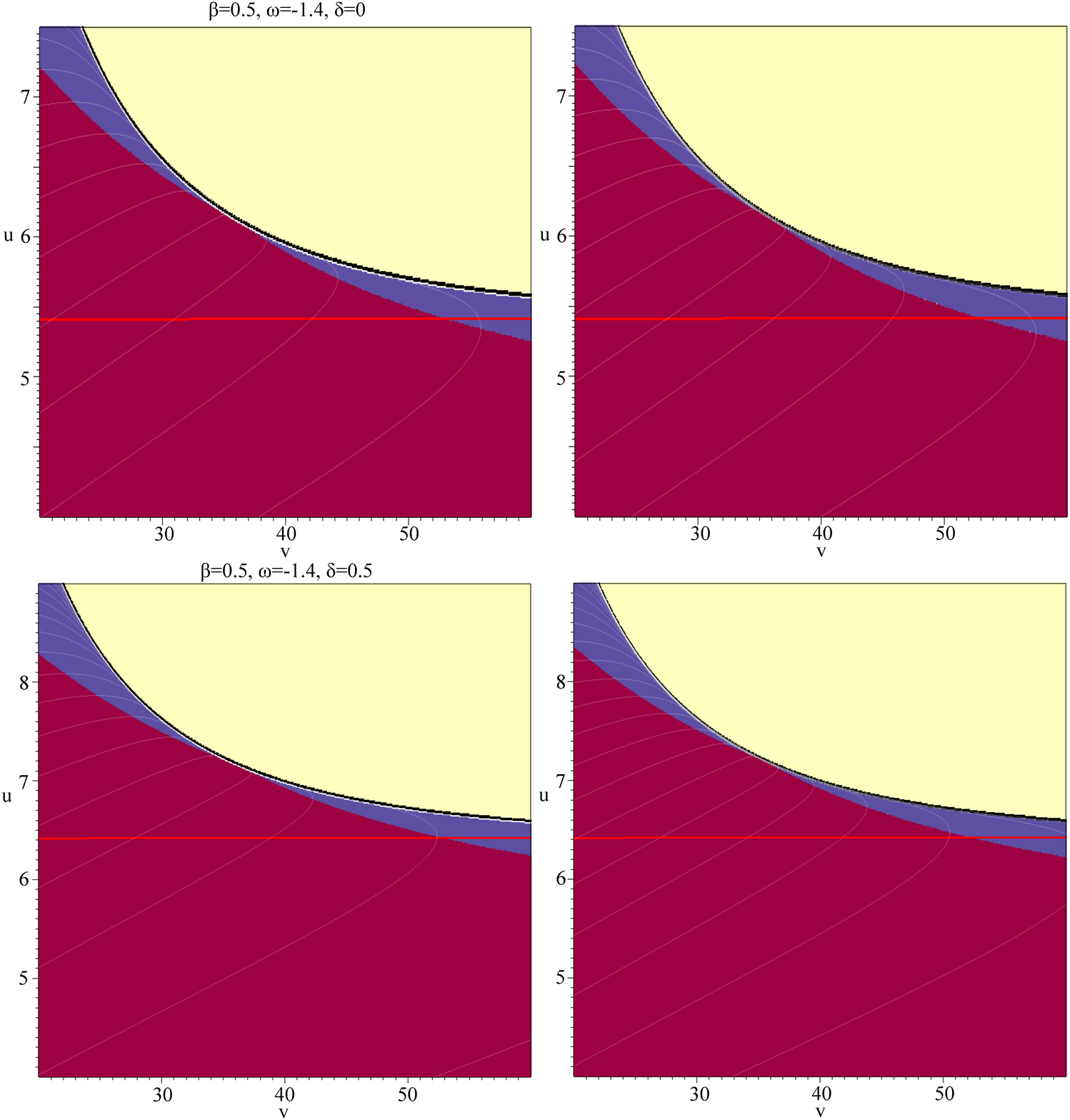}
\caption{(color online) Field contours in~the~asymptotic region of~spacetimes formed in~the~collapse within the~Brans-Dicke theory for $\beta$ equal to~$0.5$. Left column: the~Brans-Dicke field. Right column: the~real part of~the~complex scalar field. The~steps between the~contours are $\Delta \Phi = 0.0025$ and~$\Delta \phi_{Re} = 0.002$.}
\label{fig:beta05-enla}
\end{figure}

Figures~\ref{fig:beta0-BD} and~\ref{fig:beta0-re} present the~hypersurfaces of~constant Brans-Dicke field and~the~scalar field, respectively, in~spacetimes formed during evolutions conducted for $\beta=0$. The~enlarged dynamical and~asymptotic regions of~the~vicinity of~the~central singularity are shown in~figure~\ref{fig:beta0-enla}. The~constant Brans-Dicke field and~scalar field hypersurfaces in~the~spacetimes which stem from the~gravitational collapse for $\beta=0.5$ are shown in~figures~\ref{fig:beta05-BD} and~\ref{fig:beta05-re}, respectively. The~selected asymptotic regions neighboring the~central singularity were enlarged and~are depicted in~figure~\ref{fig:beta05-enla}.

The~dynamics of~the~Brans-Dicke field depends on~both parameters $\omega$ and~$\delta$ in~the~cases of~$\beta$ equal to~$0$ and~$0.5$. It~is especially clear for spacetimes obtained with $\omega\geqslant -1.4$, for which the~field varies more substantially and~at~earlier retarded times as~$\omega$ decreases. The~most considerable field dynamics is observed at~late retarded times nearby the~singularity. In~the~ghost limit, i.e.,~for $\omega=-1.6$, the~field function varies extensively in~the~vicinity of~the~Cauchy horizon, which is the~limit of~predictability of~the~evolution equations. The~influence of~the~parameter~$\delta$ on~the~Brans-Dicke field dynamics is far less apparent in~comparison to~$\omega$. In~the~case of~the~evolution proceeding in~the~presence of~the~real scalar field, the~variations of~the~Brans-Dicke field function are slightly smaller in~comparison to~the~ones observed during evolutions with $\delta=0.5$.

A~spacelike character of~the~hypersurfaces along which the~Brans-Dicke field is constant is manifested in~the~regions of~high curvature, that is nearby the~central singularity, in~all dynamical spacetimes. There exist separated points, at~which the~$r=0$ singularity seems to~have one common point with the~regions of~spacelike lines of~constant Brans-Dicke field. In~the~case of~$\delta=0$, such points exist along the~whole central singularity, while for the~complex scalar field they appear only in~the~asymptotic region, that is for large values of~advanced time. It~means that the~presence of~a~complex scalar field is conducive to~measuring time with the~use of~the~accompanying Brans-Dicke field, especially in~the~dynamical spacetime region. The~exact determination whether in~these points single null or~timelike constancy hypersurfaces reach the~singularity is impossible because of~the~limitations associated with conducting numerical calculations nearby the~singularity. However, even if such isolated points do exist, they do not influence the~usefulness of~the~Brans-Dicke field in~time quantification in~the~vicinity of~the~singularity. A~similar situation was also observed for a~gravitationally collapsing single real scalar field in~Einstein gravity~\cite{NakoniecznaLewandowski2015-064031}. The~lines indicating equal values of~the~Brans-Dicke field function in~the~spacetime are timelike in~the~vicinity of~the~Cauchy horizon, which forms when $\omega$ equals $-1.6$, so in~this spacetime region the~field definitely cannot be used as~a~time measurer. It~should be emphasized that a~potential time measuring with the~use of~the~Brans-Dicke field in~the~considered theoretical setup nearby the~singularity is possible also due to~the~fact that the~field constancy hypersurfaces vary monotonically in~this area.

The~behavior of~the~hypersurfaces of~constant values of~the~scalar field function is uniform for $\omega\geqslant -1.4$ with only minor differences for the~two investigated values of~the~parameter $\delta$. The~scalar field is certainly more dynamical in~the~case of~$\omega=-1.6$ in~the~neighborhood of~the~Cauchy horizon and~the~central singularity at~late retarded times. The~field constancy hypersurfaces are spacelike in~the~vicinity of~the~singularity and~only isolated points at~the~singularity in~which they may potentially be non-spacelike exist, similarly to~the~case of~the~Brans-Dicke field described above. The~timelike character of~the~hypersurfaces is observed nearby the~Cauchy horizon and~for this reason the~scalar field cannot serve as~a~time measurer there. The~changes of~the~field function values are monotonic as~the~singularity is approached, which is also a~necessary condition in~order to~quantify time using the~evolving scalar field.

\subsection{Heterotic model}
\label{sec:fields-het}

\begin{figure}[tbp]
\centering
\includegraphics[width=0.8\textwidth]{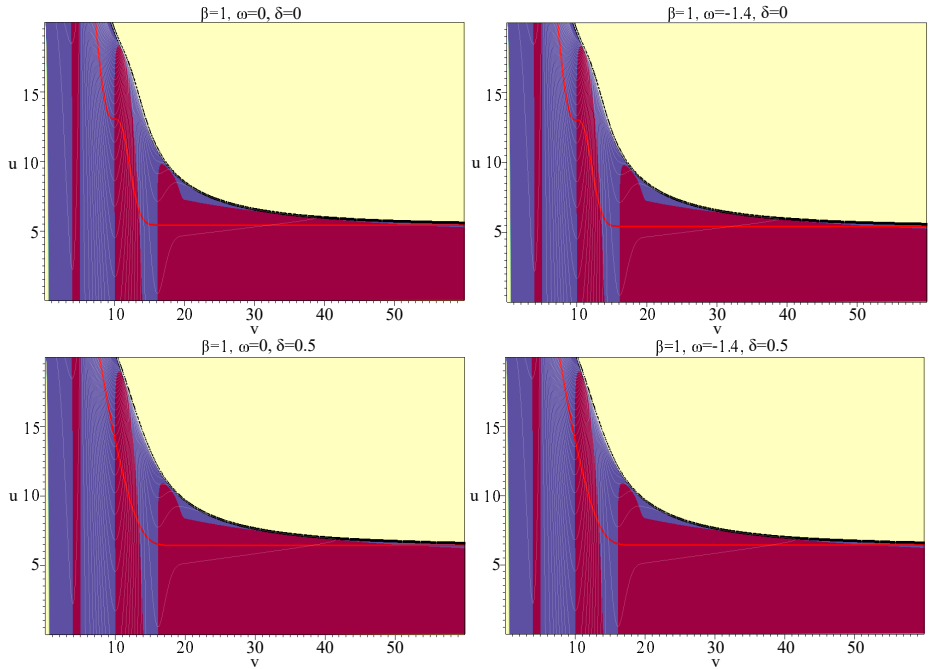}
\caption{(color online) The~contours of~the~real part of~the~complex scalar field for evolutions conducted within the~Brans-Dicke theory for $\beta=1$.}
\label{fig:beta1-re}
\end{figure}

When $\beta$ equals $1$, the~evolution equation of~the~Brans-Dicke field~\eqref{eqn:BDfield-eqn2} reduces to~the~homogeneous wave equation written in~double null coordinates. Thus, the~initially constant field function remains unchanged within the~whole computational domain. For~this reason, the~Brans-Dicke field cannot be used as~a~time variable in~this case. The~hypersurfaces of~the~constant scalar field function in~the~case of~$\beta=1$ are presented in~figure~\ref{fig:beta1-re}. It~turns out that the~scalar field dynamics does not depend on~the~evolution parameters. As~in~the~previous cases, the~monotonicity of~the~field function is observed in~the~region of~high curvature near the~central singularity. The~field constancy hypersurfaces are also spacelike in~this region, apart from single separated points, in~which a~timelike or~null hypersurface can potentially reach the~central $r=0$ singular line. Due to~the~above arguments, although the~Brans-Dicke field is excluded as~a~clock in~the~heterotic theory, the~scalar field remains a~good candidate in~this regard.

\subsection{Overall dependence on~the~evolution parameters}
\label{sec:fields-sum}

When the~influence of~the~$\beta$ coupling constant on~the~overall field dynamics during the~collapse is considered, it~turns out that there are no conspicuous differences between the~process proceeding within the~type IIA~and type I models. Significant distinctness is observed for $\beta=1$, i.e.,~in~the~heterotic theory, mainly due to~the~fact that the~equation of~motion of~the~Brans-Dicke field~\eqref{eqn:BDfield-eqn2} simplifies as~was described above and~the~field dynamics becomes trivial. The~parameter $\omega$ strongly influences the~behavior of~the~Brans-Dicke field, whose dynamics is most significant for~$\omega$ close to~$-1.5$, while its impact on~the scalar field dynamics is unnoticeable.

\section{Conclusions}
\label{sec:conclusions}

The~dynamical gravitational collapse of~complex and~real scalar fields in~the~Brans-Dicke theory was investigated. The~structures of~the~emerging spacetimes were examined and the~feasibility of~performing time measurements with the~use of~evolving scalar fields during dynamical processes driven by the~gravitational interaction was assessed. Several values of~the Brans-Dicke coupling constant, which corresponded to~the~large~$\omega$, $f(R)$, dilatonic, brane-world and~ghost limits, were investigated. The~studied coupling between the~Brans-Dicke field and~the~matter sector of~the theory, which was controlled by the~$\beta$ parameter, was motivated by the~type IIA, type I and~heterotic string theory-inspired models.

In~the~case of~$\omega$ equal to~$10$, $0$ and~$-1$ in~the~type IIa and~type I models, as~well as~for all its values within the~heterotic model, in~the~spacetimes which stem from the~collapse of~both real and~complex scalar fields there exists a~spacelike central singularity surrounded by a~single apparent horizon. When $\beta=0$ and~the~Brans-Dicke coupling equals $-1.4$, an~additional horizon appears in~the~spacetime at~late retarded times. It~is absent in~the~case of~$\beta=0.5$ and~$\omega=-1.4$, $\delta=0.5$. During the~collapse of~a~real scalar field with $\beta=0$ and~$\omega=-1.6$, two parts of~a~spacelike central singularity surrounded separately by the~apparent horizons form. They are linked by a~Cauchy horizon null segment, which is visible for a~distant observer. During the~collapse of~a~complex scalar field, the~emerging spacetime contains a~spacelike singularity along $r=0$, which is situated beyond a~single apparent horizon. In~all the~cases, the~apparent horizon settles along the~$u=const.$ hypersurface as~$v\to\infty$, i.e.,~in~the~non-dynamical part of~the spacetime.

When $\beta$ equals $0$ and~$0.5$, the~Brans-Dicke field dynamics is more considerable at~earlier retarded times as~$\omega$ decreases. The~values of~the field function vary most significantly in~the~vicinity of~the central singularity and~nearby the~Cauchy horizon, if it~exists in~the~spacetime. The~variations of~the~Brans-Dicke field function are slightly smaller when it~is accompanied by the~real scalar field, when compared with the~evolution proceeding in~the~presence of~a~complex scalar field. The~case of~$\beta=1$ excludes the~Brans-Dicke field from being treated as~a~time variable, because due to~the~form of~its evolution equation, it~remains constant within the~whole dynamical spacetime.

In~all dynamical spacetimes obtained in~type IIa and~type I models the~constancy hypersurfaces of~the Brans-Dicke field are spacelike nearby the~central singularity. There are several points, at~which a~non-spacelike hypersurface can potentially reach the~singularity. However, due to~the~fact that these points are separated, they do not prevent the~field from acting as~a~time measurer. The~constancy hypersurfaces of~the Brans-Dicke field are timelike nearby the~Cauchy horizon and~hence they cannot serve as~`clocks' there. The~potential usefulness of~the~Brans-Dicke field for time measurements nearby the~singularity is possible also due to~the~fact that the~field constancy hypersurfaces vary monotonically in~this area.

In~the~cases of~$\beta$ equal to~$0$ and~$0.5$, the~behavior of~the~scalar field function constancy hypersurfaces displays only minor differences for the~two investigated values of~the~parameter $\delta$. The~scalar field dynamics is most noticeable in~the~neighborhood of~the~Cauchy horizon and~the~central singularity at~late retarded times. The~hypersurfaces of~constant field function are spacelike in~the~vicinity of~the~singularity and~only isolated points at~the~singularity in~which they may potentially be non-spacelike exist. The~timelike character of~the~hypersurfaces is observed nearby the~Cauchy horizon and~for this reason the~scalar field cannot serve as~a~time measurer in~this area. The~changes of~the~field function values are monotonic as~the~singularity is approached. In~the~case of~$\beta=1$, the~scalar field dynamics does not depend on~the~evolution parameters and~the~above qualitative description applies also to~this case. Although the~Brans-Dicke field is excluded as~a~clock in~the~heterotic theory, the~scalar field remains a~good candidate in~this regard.

In conclusion, using scalar fields as~time variables within the~whole evolving spacetime during dynamical gravitational evolutions of~coupled matter-geometry systems encounters several obstacles, which can be problematic and~should be remembered. First, the~two conditions which are necessary for treating the~field as~a~time measurer (spacelike character of~its constancy hypersurfaces and~monotonicity of~their parametrization) are not fulfilled in~the whole spacetime. Second, the~vicinity of~Cauchy horizons should be excluded from such analyses, at~least within the~studied theoretical setup. Third, the~forms of~the~field evolution equations should be analyzed thoroughly for various values of~parameters which they contain, because the~possibility of~using the~specific scalar field as~a~time variable may be excluded in~some cases. Fortunately, only the~last of~the~outlined difficulties applies to~a~close proximity of~the~singularity emerging in~the spacetime. This region of~high curvature is of~crucial importance from the~viewpoint of~gravity quantization, which is the~main reason of~the~undertaken studies. Hence, the~scalar fields can be used to~quantify time nearby the~spacetime singularity (provided that the~equation of~motion of~the~particular field does not reduce to~the~wave equation and~thus the~values of~the~field functions vary there).

The~investigated case was the~neutral gravitational collapse. There exists a~question whether additional gauge vector field can influence the~obtained results and~either strengthen or weaken the~conclusion that both real and~complex scalar fields can be used as~time variables during dynamical evolutions of~coupled matter-geometry systems. Since the~studied process proceeding in~the presence of~an~electric charge is a~toy-model for the~realistic collapse~\cite{HodPiran1998-1554,HodPiran1998-1555,SorkinPiran2001-084006,SorkinPiran2001-124024,OrenPiran2003-044013}, we plan to~address this issue in~the future researches. It~will also allow us to~investigate the~regions neighboring Cauchy horizons in~more detail, as~these null hypersurfaces appear naturally during the~charged collapse.

\appendix
\section{Numerical computations}
\label{sec:appendix}

The~scheme employed in~the~numerical simulations was described in~detail in~\cite{HansenKhokhlovNovikov2005-044013,DoroshkevichHansenNovikovNovikovParkShatskiy2010-124011,HongHwangStewartYeom2010-045014,HwangYeom2011-064020,HansenLeeParkYeom2013-235022,HwangYeom2010-205002,HansenYeom2014-040,HansenYeom2015-019}. The~initial conditions were imposed on~a~set of~dynamical variables ($\alpha$, $h$, $d$, $r$, $f$, $g$, $\Phi$, $W$, $Z$, $s$, $w$, $z$) on~two null $u=const.$ and~$v=const.$ hypersurfaces, which were assumed to~be $u=0$ and~$v=0$ for the~need of~the numerical setup. The~gauge freedom of~the $r$ function was fixed by imposing constant $r|_{(0,0)}$, $f|_{v=0}$ and~$g|_{u=0}$, where the~last two were negative and~positive, respectively, so that the~radial function decreased for an~ingoing observer and~increased for an~outgoing one. The~values of~$Z$ and~$z$ along $u=0$ stemmed from initial conditions~\eqref{eqn:field-ini} and~\eqref{eqn:BDfield-ini}. At the~hypersurface $v=0$, the~Brans-Dicke field was set as~equal to~unity and, due to~the~shell-shaped form of~the scalar field, the~function $s$ was constant and~equal to~$A$. The~functions $W$ and~$w$ were thus also specified. The~metric function $\alpha$ was equal to~$1$ which gave $h=0$ at~$v=0$, since this axis was not affected by the~evolving scalar field in~the form of~a~shell. The~remaining initial conditions were calculated with the~use of~the above foundations and~appropriate equations from among the~set~\eqref{eqn:matrix-2}--\eqref{eqn:field-eqn-uv}.

\begin{figure}[tbp]
\centering
\includegraphics[width=0.75\textwidth]{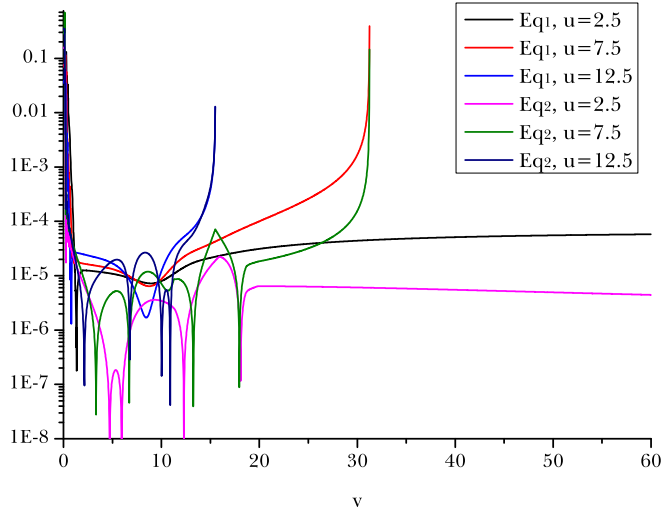}
\vspace{-0.35cm}
\caption{(color online) Monitoring of~the constraints. The~values of~the equations~\eqref{eqn:constr1} and~\eqref{eqn:constr2} were calculated along three null hypersurfaces of~constant~$u$ equal to~$2.5$, $7.5$ and~$12.5$.}
\label{fig:const}
\end{figure}

The~correctness of~the numerical code was checked using a~sample evolution described by parameters $\beta=0$, $\omega=-1$ and~$\delta=0.5$. The~constraints were monitored during the~process using the~following equations:
\ben\label{eqn:constr1}
\textrm{Eq}_1 &\equiv& \frac{r_{,uu} - 2fh + 4 \pi r \left( T_{uu}^{\Phi} + T_{uu}^{\mathrm{M}}\Phi^{-1} \right)}{|r_{,uu}| + |2fh| + 4\pi r \left( \left| T_{uu}^\Phi \right| + \left| T_{uu}^{\mathrm{M}}\Phi^{-1} \right| \right)}, \\
\label{eqn:constr2}
\textrm{Eq}_2 &\equiv& \frac{r_{,vv} - 2gd + 4 \pi r \left( T_{vv}^{\Phi} + T_{vv}^{\mathrm{M}}\Phi^{-1} \right)}{|r_{,vv}| + |2gd| + 4\pi r \left( \left| T_{vv}^\Phi \right| + \left| T_{vv}^{\mathrm{M}}\Phi^{-1} \right| \right)}.
\een
The~above relations as~functions of~advanced time calculated for three selected values of~retarded time are shown in~figure~\ref{fig:const}. In~both cases these should be zero in~principle and~the~deviations appear due to~numerical errors. There are two regions in~which the~values rise considerably, i.e.,~for small values of~$v$ due to~the~fact that the~denominators are very close to~zero and~nearby the~singularity. As~can be inferred from the~plot, except a~close neighborhood of~the~singularity, the~errors do not exceed $0.1\%$. Since the~constraint equations are stable, the~simulations are consistent.

\begin{figure}[tbp]
\centering
\includegraphics[width=0.75\textwidth]{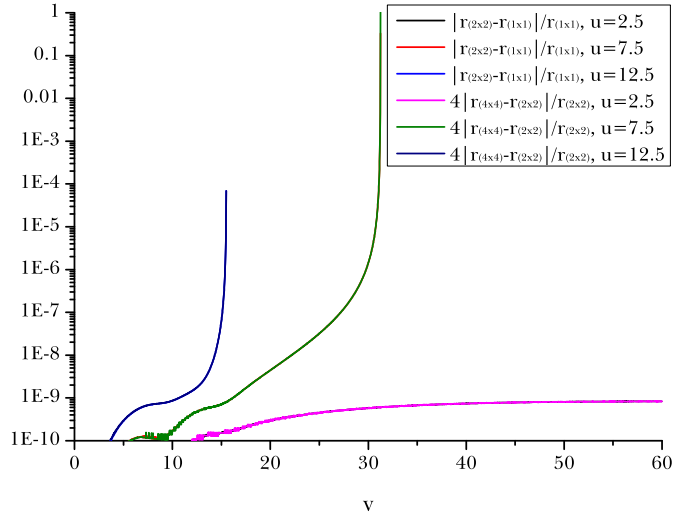}
\vspace{-0.35cm}
\caption{(color online) The~convergence of~the~code presented through the~prism of~the~values of~the~quantity $|r_{(2k\times 2k)}-r_{(k\times k)}| /r_{(k\times k)}$ with $k=1,2$ calculated at~the~same hypersurfaces of~constant~$u$ as~in~figure~\ref{fig:const}. $(k\times k)$ denotes the~resolution of~the~numerical grid, on~which the~computations were conducted.
}
\label{fig:conv}
\end{figure}

The~convergence of~the code was checked and~the outcome is presented in~figure~\ref{fig:conv}. The~values of~a~quantity constructed from the~$r$ function obtained on~two grids with a~quotient of~integration steps equal to~$2$ were calculated along three arbitrarily chosen $u=const.$ lines. An~overlap between two profiles of~the defined quantity at~each $u=const.$ was obtained when the~result from finer grids was multiplied by $4$. Thus, the~code displays a~second order convergence. The~discrepancy between each two profiles at~the constant~$u$ is less than $0.01\%$ except a~close vicinity of~the~singularity. Hence, the~coarsest grid was adequate for performing the~simulations.

\acknowledgments

A.N. would like to~thank Professor Jerzy Lewandowski for~drawing her attention to~the~problem considered in~the~article and~encouraging discussions. A.N. was partially supported by~the~Polish National Science Centre grant no.~DEC-2014/15/B/ST2/00089. D.Y. is supported by~Leung Center for~Cosmology and~Particle Astrophysics (LeCosPA) of~National Taiwan University (103R4000).





\bibliographystyle{JHEP}
\bibliography{measuringtime.bib}

\end{document}